\newcommand{\hsizeforfig}{5in}
\newcommand{\hsizeforfigthreequarters}{3.75in}
\newcommand{\hsizeforfighalf}{2.5in}

\documentclass{ws-ijmpe}
\usepackage[super,compress]{cite}

\usepackage{graphicx}
\usepackage{amssymb}
\usepackage{isotope}
\usepackage{mathrsfs}
\usepackage{subscript}

\usepackage{liealg}
\usepackage{wignert}

\usepackage{mciop}  
\renewcommand{\tfrac}{\textstyle\frac}  

\newcommand{\Nmax}{{N_\text{max}}}
\newcommand{\scrD}{{\mathscr{D}}}

\newcommand{\scrR}{{\mathscr{R}}}
\newcommand{\MeV}{{\mathrm{MeV}}}
\newcommand{\keV}{{\mathrm{keV}}}

\newcommand{\hw}{{\hbar\omega}}

\newcommand{\Jvec}{\vec{J}}
\newcommand{\Lvec}{\vec{L}}
\newcommand{\Svec}{\vec{S}}

\newcommand{\Qvec}{\vec{Q}}
\newcommand{\Mvec}{\vec{M}}

\newcommand{\Dvec}{\vec{D}}

\newcommand{\glp}{{g_{\ell,p}}}
\newcommand{\gln}{{g_{\ell,n}}}
\newcommand{\gsp}{{g_{s,p}}}
\newcommand{\gsn}{{g_{s,n}}}

\newcommand{\Dlp}{{\Dvec_{\ell,p}}}
\newcommand{\Dln}{{\Dvec_{\ell,n}}}

\newcommand{\nnloopt}{NNLO\textsubscript{opt}}
\newcommand{\nthreelo}{N\textsuperscript{3}LO}

\begin{document}

\markboth{M.~A.~Caprio, P.~Maris, J.~P.~Vary  \& R.~Smith}{Collective rotation from \textit{ab initio} theory}

\catchline{}{}{}{}{}

\title{Collective rotation from \textit{ab initio} theory}

\author{M.~A.~Caprio}

\address{Department of Physics, University of Notre Dame,\\ Notre Dame, Indiana 46556-5670, USA}

\author{P.~Maris and J.~P.~Vary}

\address{Department of Physics and Astronomy, Iowa State University,\\ Ames, Iowa 50011-3160, USA}

\author{R.~Smith}

\address{School of Physics and Astronomy, University of Birmingham,\\ Edgbaston, Birmingham, B15 2TT, UK}

\maketitle

\begin{history}
\end{history}

\begin{abstract}
Through \textit{ab initio} approaches in nuclear theory, we may now
seek to quantitatively understand the wealth of nuclear collective
phenomena starting from the underlying internucleon interactions.
No-core configuration interaction (NCCI) calculations for $p$-shell
nuclei give rise to rotational bands, as evidenced by rotational
patterns for excitation energies, electromagnetic moments, and
electromagnetic transitions.  In this review, NCCI calculations of
$\isotope[7\text{--}9]{Be}$ are used to illustrate and explore
\textit{ab initio} rotational structure, and the resulting predictions
for rotational band properties are compared with experiment.  We
highlight the robustness of \textit{ab initio} rotational predictions
across different choices for the internucleon interaction.
\end{abstract}


\ccode{PACS numbers: 21.60.Cs, 21.10.-k, 21.10.Re, 27.20.+n}


\section{Introduction}
\label{sec-intro}

The challenge of \textit{ab initio} nuclear theory is to
quantitatively predict the complex and highly-correlated behavior of
the nuclear many-body system, starting from the underlying
internucleon interactions.  Significant progress has been made in the
$\textit{ab initio}$ description of light nuclei through large-scale
calculations.~\cite{pieper2004:gfmc-a6-8,neff2004:cluster-fmd,hagen2007:coupled-cluster-benchmark,bacca2012:6he-hyperspherical,shimizu2012:mcsm,barrett2013:ncsm,shirokov2014:jisp16-binding-SPECIAL}
We may now seek to understand the wealth of nuclear collective
phenomena~\cite{rowe2010:collective-motion} through  \textit{ab
  initio}
approaches~\cite{wiringa2000:gfmc-a8,dytrych2007:sp-ncsm-dominance,neff2008:clustering-nuclei,shimizu2012:mcsm,dytrych2013:su3ncsm}.

In particular, rotational bands emerge in \textit{ab initio} no-core
configuration interaction (NCCI)~\cite{barrett2013:ncsm} calculations of $p$-shell
nuclei~\cite{caprio2013:berotor,maris2015:berotor2}.
Rotational patterns are found in the calculated level energies,
electromagnetic moments, and electromagnetic transitions.  
Natural questions
surrounding the emergence of rotation in \textit{ab initio}
calculations include:
\begin{romanlist}[(ii)]
\item How recognizable is
  the rotation, from the calculated observables?
\item How robust
  is the prediction of rotation, both against limitations in the
  many-body calculation and, more fundamentally, against uncertainties in the 
  internucleon interaction?
\item How, physically, does the rotation arise, or what is the
  intrinsic structure?
\item How well does the calculated rotation  agree with
  experiment, when compared quantitatively?
\end{romanlist}
However, to understand the emergence of rotation in NCCI calculations
and address these questions, we must first consider the
\textit{ab initio} calculations themselves.  NCCI
calculations are, of necessity, carried out in
a finite, truncated space.  Computational restrictions limit the
extent to which converged calculations can be obtained.

This review is based upon the ideas and results of recent analyses of
rotation in \textit{ab initio} NCCI calculations.  A systematic study
of the emergence of rotational bands in NCCI calculations of
$\isotope[7-12]{Be}$, using the JISP16 nucleon-nucleon
interaction~\cite{shirokov2007:nn-jisp16}, is presented in
Refs.~\refcite{caprio2013:berotor,maris2015:berotor2,caprio2015:berotor-brasov14}.
The spin and orbital angular momentum structure of rotational states
in $\isotope[7]{Li}$ (the mirror nucleus to $\isotope[7]{Be}$) and
$\isotope[9]{Be}$ is investigated, using a chiral next-to-next-to-next-to-leading-order (\nthreelo{})
interaction~\cite{entem2003:chiral-nn-potl}, in
Ref.~\refcite{johnson2015:spin-orbit}.  (NCCI calculations for the
ground state rotational band in $\isotope[12]{C}$, although not
considered in this review, are discussed in
Refs.~\refcite{maris2012:mfdn-hites12,maris2013:ncci-chiral-ccp12,maris2015:ncsm-12c-chiral}.)
We highlight here the robustness of the \textit{ab initio} rotational
predictions across different choices for the internucleon interaction.
In particular, in many of the illustrations, we compare the results of
calculations based on two interactions obtained by very different
procedures: the JISP16 interaction (mentioned above) and the 
chiral next-to-next-to-leading-order (NNLO) interaction
\nnloopt{}~\cite{ekstroem2013:nnlo-opt}.

The approach of this review is not to attempt an exhaustive summary of
the rotational phenomena noted in recent NCCI calculations, but rather
to focus on exploring a few illustrative cases.  Specifically,
calculations of $\isotope[7\text{--}9]{Be}$ are used to illustrate
emergent rotational phenomena and to exemplify some of the ideas
involved in analysis of \textit{ab initio} rotational structure.  We
begin by introducing the challenges in obtaining converged results for
the relevant observables in \textit{ab initio} calculations
(Sec.~\ref{sec-ncci}).  The definition of rotation in nuclei and its
expected signatures are then briefly reviewed (Sec.~\ref{sec-rot}).
Successively richer examples of rotation in the $\isotope{Be}$
isotopes are examined in Sec.~\ref{sec-be}: the even-even isotope
$\isotope[8]{Be}$ (Sec.~\ref{sec-be-8be}), the odd-mass isotope
$\isotope[7]{Be}$ (Sec.~\ref{sec-be-7be}), and rotational structure
including excited bands (and both parities) in $\isotope[9]{Be}$
(Sec.~\ref{sec-be-9be}).  Finally, we compare the rotational energy
parameters extracted from \textit{ab initio} calculations with those
for the experimentally observed bands in $\isotope[7\text{--}9]{Be}$
(Sec.~\ref{sec-be-params}).


\section{NCCI calculations and their convergence}
\label{sec-ncci}
\begin{figure}[t]
\centerline{\includegraphics[width=\hsizeforfigthreequarters]{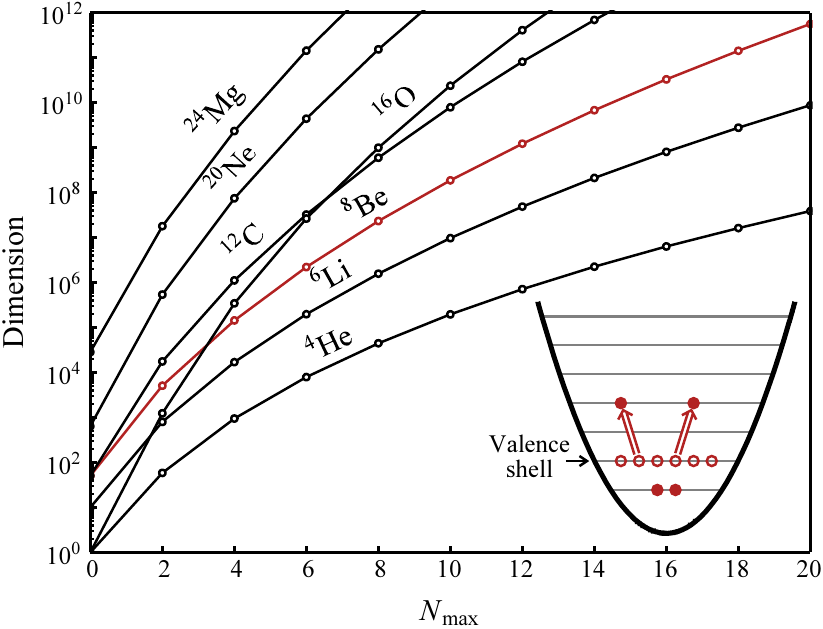}}
\caption{Dimensions for NCCI calculations, as a function of the number
  of oscillator excitations $\Nmax$ included in the basis, for
  selected nuclides.  
  The example configuration shown~(inset) involves a total of four oscillator
  excitations above the lowest oscillator configuration and thus would
  be included in calculations with $\Nmax=4$ and higher (for
  simplicity, only single particle states for one type of nucleon,
  protons or neutrons, are shown).  Dimensions are shown for $M$-scheme
  natural parity, $M=0$ spaces (see Sec.~\ref{sec-be-8be} text).  }
\label{fig-ncci-dim}      
\end{figure}

In NCCI calculations, the nuclear many-body Schr\"odinger equation is
formulated as a Hamiltonian matrix eigenproblem.  The Hamiltonian is
represented with respect to a basis of antisymmetrized products of
single-particle states.  Conventionally, harmonic oscillator
states~\cite{moshinsky1996:oscillator} are used as the single-particle
states, for the technical convenience they provide (both in
transforming interaction matrix elements between relative and
single-particle coordinates and in
obtaining an exact separation of the center-of-mass wave
function).  The problem is then solved for the
full system of $A$ nucleons, \textit{i.e.}, with no inert core.  

In
practice, calculations must be carried out in a finite-dimensional
subspace, commonly obtained by truncating the basis to a maximum
allowed number $\Nmax$ of oscillator
excitations.  Convergence
toward the exact results~--- as would be achieved in the full,
infinite-dimensional space~--- is obtained with increasing $\Nmax$.
However, the basis size grows combinatorially with $\Nmax$, so the
maximum accessible $\Nmax$ is severely limited by computational
restrictions.  The dimensions for representative cases are shown in
Fig.~\ref{fig-ncci-dim} (and the meaning of $\Nmax$ is illustrated in
the inset).  Thus, \textit{e.g.}, $\Nmax=10$ calculations for
$\isotope[8]{Be}$, as considered below, involve a Hamiltonian matrix
dimension of $\sim2\times10^8$.

The calculated eigenvalues and wave functions, and thus the calculated
values for observables,
depend both upon the basis truncation $\Nmax$ and on the length
parameter $b$ for the oscillator basis functions, which is customarily
specified by the equivalent oscillator energy
$\hw$~\cite{suhonen2007:nucleons-nucleus}.  Any attempt to interpret the results of NCCI
calculations (Sec.~\ref{sec-be}) or compare the
calculations with experiment (Sec.~\ref{sec-be-params}) must take into account  the manner in
which these observables approach convergence and the
level of convergence which has been achieved.
\begin{figure}[t]
\centerline{\includegraphics[width=\hsizeforfig]{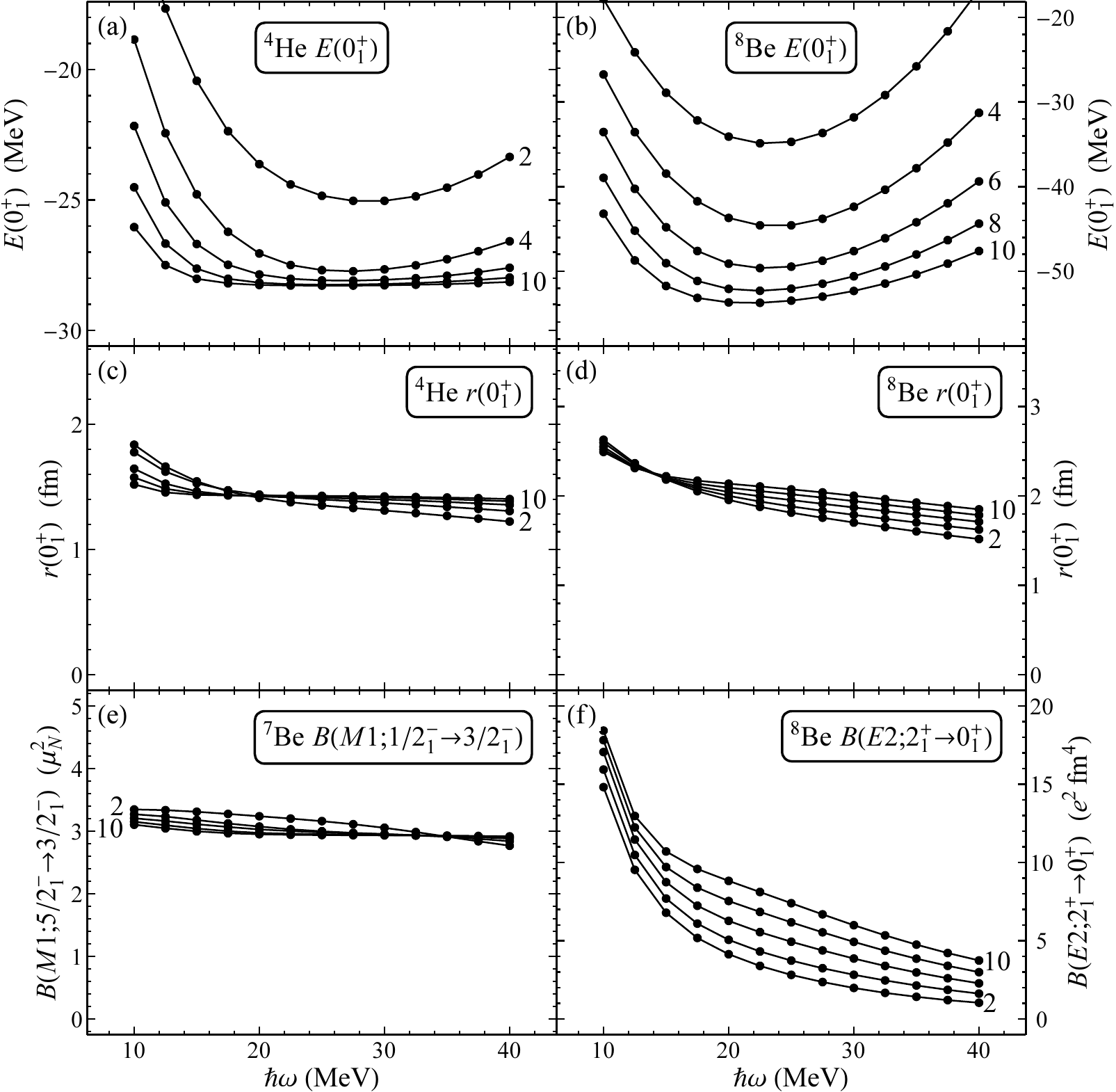}}
\caption{The
  $\Nmax$ and $\hw$ dependence of values obtained for observables in
  NCCI calculations, including comparatively converged and unconverged
  cases: (a)~the $\isotope[4]{He}$ ground state energy eigenvalue, (b)~the
  $\isotope[8]{Be}$ ground state energy eigenvalue, (c)~the
  $\isotope[4]{He}$ ground state RMS matter radius, (d)~the
  $\isotope[8]{Be}$ ground state RMS matter radius, (e)~the magnetic dipole
   reduced transition probability between the first excited
  state and ground state of $\isotope[7]{Be}$, and (f)~the electric
  quadrupole reduced transition probability between the first excited
  state and ground state of $\isotope[8]{Be}$.  
  Calculated values are shown as functions of $\hw$ for
  $\Nmax=2$ to $10$ (as
  labeled) and are obtained with the JISP16 nucleon-nucleon
  interaction, with Coulomb interaction between protons.
}
\label{fig-conv}      
\end{figure}

The $\Nmax$ and $\hw$ dependences of calculated ground-state energy
eigenvalues are illustrated for $\isotope[4]{He}$ in
Fig.~\ref{fig-conv}(a) and for $\isotope[8]{Be}$ in
Fig.~\ref{fig-conv}(b), in both cases for $2\leq \Nmax\leq10$ (we use
the same $\Nmax$ range in all calculations, for purposes of
comparison, although current computational limits are significantly
higher for $\isotope[4]{He}$).  For a fixed $\Nmax$, a minimum in the
calculated energy is obtained at some $\hw$ (for the $\isotope{Be}$
isotopes this will typically be in the range
$\hw\approx20$--$25\,\MeV$).  By the variational principle, 
any such calculated energy in a truncated space provides an upper bound on the
true ground state energy in the full, untruncated many-body space.
As $\Nmax$ is increased, a lower calculated ground state energy
is obtained at each $\hw$.  The approach to convergence is marked by
approximate $\Nmax$ independence (a compression of successive energy
curves) and $\hw$ independence (a flattening of each curve around its
minimum).  While a high level of convergence (at the $\keV$ scale) may
be obtained in the lightest nuclei, in particular, for the tightly
bound and compact nucleus $\isotope[4]{He}$ [Fig.~\ref{fig-conv}(a)],
the situation is more challenging for the $\isotope{Be}$ isotopes.
The decrease in the variational minimum energy for $\isotope[8]{Be}$
does become smaller with each step in $\Nmax$
[Fig.~\ref{fig-conv}(b)], but even at $\Nmax=10$ these changes are
still at the $\MeV$ scale.

For electric quadrupole moments and transition strengths,
traditionally so important in the identification of rotational
structure~\cite{bohr1998:v2}, convergence is even more elusive.  The
quadrupole operator, which has the form $Q_{2,m}\propto r^2
Y_{2,m}$~\cite{suhonen2007:nucleons-nucleus}, includes an $r^2$ radial
dependence and is therefore highly sensitive to the large-$r$ ``tails'' of the
nuclear wave function, which are poorly reproduced in a harmonic
oscillator basis (see, \textit{e.g.}, Fig.~1 of Ref.~\refcite{caprio2014:cshalo}). 

While it is difficult to come by an illustration of successful
convergence of an electric quadrupole strength in NCCI calculations,
the convergence of the root mean square (RMS) radius observable in
$\isotope[4]{He}$, shown in Fig.~\ref{fig-conv}(c), provides a model
of the behavior which might be expected.  The RMS radius, like
quadrupole observables, is deduced from matrix elements of an operator
with an $r^2$ dependence.  Convergence~--- in general, manifested in
$\Nmax$ independence and $\hw$ independence~--- is here reflected in a
compression of successive $\Nmax$ curves and a flat ``shoulder'' in
the plot of the $B(E2)$ against $\hw$, over some range of $\hw$
values.  The calculated RMS radius of $\isotope[8]{Be}$, shown for
comparison in Fig.~\ref{fig-conv}(d), appears to be approaching
convergence but is not fully converged.  

Returning, finally, to the quadrupole observables, the calculated
quadrupole transition strength between the $2^+$ first excited state
and $0^+$ ground state of $\isotope[8]{Be}$ is shown in
Fig.~\ref{fig-conv}(f).  Here, the variation with $\Nmax$ and $\hw$ is
much greater,\footnote{The greater $\Nmax$ and $\hw$ dependence of the $B(E2)$
  observable [Fig.~\ref{fig-conv}(f)], as compared to the radius
  [Fig.~\ref{fig-conv}(d)], is in part an artifact of the definition
  of the observable, rather than entirely reflecting a difference in
  the actual convergence properties of the underlying matrix element.
  The RMS radius is obtained by taking the \textit{square root} of the
  expectation value [$\propto\tbracket{r^2}^{1/2}$], reducing any
  sensitivity to the matrix element, while the $B(E2)$ is obtained by
  taking the \textit{square} of the matrix element
  [$\propto\tbracket{r^2Y_2}^{2}$], amplifying any sensitivity to the
  matrix element.  Roughly speaking, the total power difference
  of $4$ in scaling with the radius between these two observables would be
  expected to quadruple all relative (percentage) sensitivities.}
and at most hints of the onset of convergence might be
apparent.
Consequently, there is no obvious way to extract quadrupole
observables, at least in their absolute magnitudes.  We shall see
(Sec.~\ref{sec-be}) that relative values of different quadrupole
observables within the same calculation may, nonetheless, be
meaningfully considered.

Magnetic dipole moments and transition strengths, in contrast, are
comparatively well-converged.  The $\Nmax$ and $\hw$ dependence of the
calculated dipole transition strength between the $1/2^-$ first excited state 
and $3/2^-$ ground state of $\isotope[7]{Be}$ is shown in
Fig.~\ref{fig-conv}(e).


\section{Collective nuclear rotation}
\label{sec-rot}

To begin, we must define what is meant by rotation in the nuclear
many-body system.  Nuclear rotation~\cite{rowe2010:collective-motion,bohr1998:v2}
arises when there is an adiabatic separation of a rotational degree of
freedom from the remaining internal degrees of freedom of the nucleus.

A rotational state factorizes into an \textit{intrinsic state}
$\tket{\phi_K}$ and a rotational wave function of the Euler angles
$\vartheta$, describing the collective rotational motion of this
intrinsic state.  Specifically, we consider an axially symmetric
intrinsic state, with definite angular momentum projection $K$ along
the intrinsic symmetry axis.  The full nuclear state
$\tket{\psi_{JKM}}$, with total angular momentum $J$ and projection
$M$, has the form
\begin{equation}
\label{eqn-psi}
\tket{\psi_{JKM}}\propto
\int d\vartheta\,\bigl[
~
\underbrace{\scrD^J_{MK}(\vartheta)}_{\text{Rotational}}
~
\underbrace{\tket{\phi_K;\vartheta}}_{\text{Intrinsic}}
~
+
~
(-)^{J+K}\scrD^J_{M,-K}(\vartheta)\tket{\phi_{\bar{K}};\vartheta}
\bigr],
\end{equation}
where $\tket{\phi_K;\vartheta}$ represents the intrinsic state
$\tket{\phi_K}$ after
rotation by $\vartheta$, and the wave function
$\scrD^J_{MK}(\vartheta)$ in the Euler angles is a Wigner $\scrD$
function.  The second term, involving the $\scrR_2$-conjugate state
$\tket{\phi_{\bar{K}};\vartheta}$, arises from discrete rotational
symmetry considerations, \textit{i.e.}, under an ``end-over-end'' rotation $\scrR_2$
by an angle $\pi$ about an axis perpendicular to the symmetry axis.  

The recognizable signatures of rotational structure reside not in the
observables for the states considered singly, but in relationships
among different rotational states arising from their closely-related
wave functions~(\ref{eqn-psi}).  A \textit{rotational band} is
comprised of nuclear states sharing the same intrinsic state
$\tket{\phi_K}$ but differing in the angular momentum $J$ of their
rotational motion, \textit{i.e.}, differing in their angular wave
functions $\scrD^J_{MK}(\vartheta)$.  Within a rotational band, $J=K$,
$K+1$, $\ldots$, except for $K=0$ bands, where only even $J$ or only
odd $J$ are present (depending upon the $\scrR_2$ symmetry).  Energies
and electromagnetic multipole matrix elements among band members
follow well-defined rotational patterns.

Band members are expected to have energies following the rotational formula
$E(J)=E_0+AJ(J+1)$, where the rotational energy constant $A\equiv\hbar^2/(2\cal{J})$ is
inversely related to the moment of inertia $\cal{J}$ of the intrinsic state.  For $K=1/2$
bands, the Coriolis contribution to the kinetic energy significantly
modifies this pattern, yielding an energy staggering
\begin{equation}
\label{eqn-EJ-stagger}
E(J)=E_0+A\bigl[J(J+1)+
\underbrace{a(-)^{J+1/2}(J+\tfrac12)}_{\text{Coriolis ($K=1/2$)}}
\bigr],
\end{equation}
where $a$ is the Coriolis decoupling parameter.

\begin{figure}
\centerline{\includegraphics[width=\hsizeforfig]{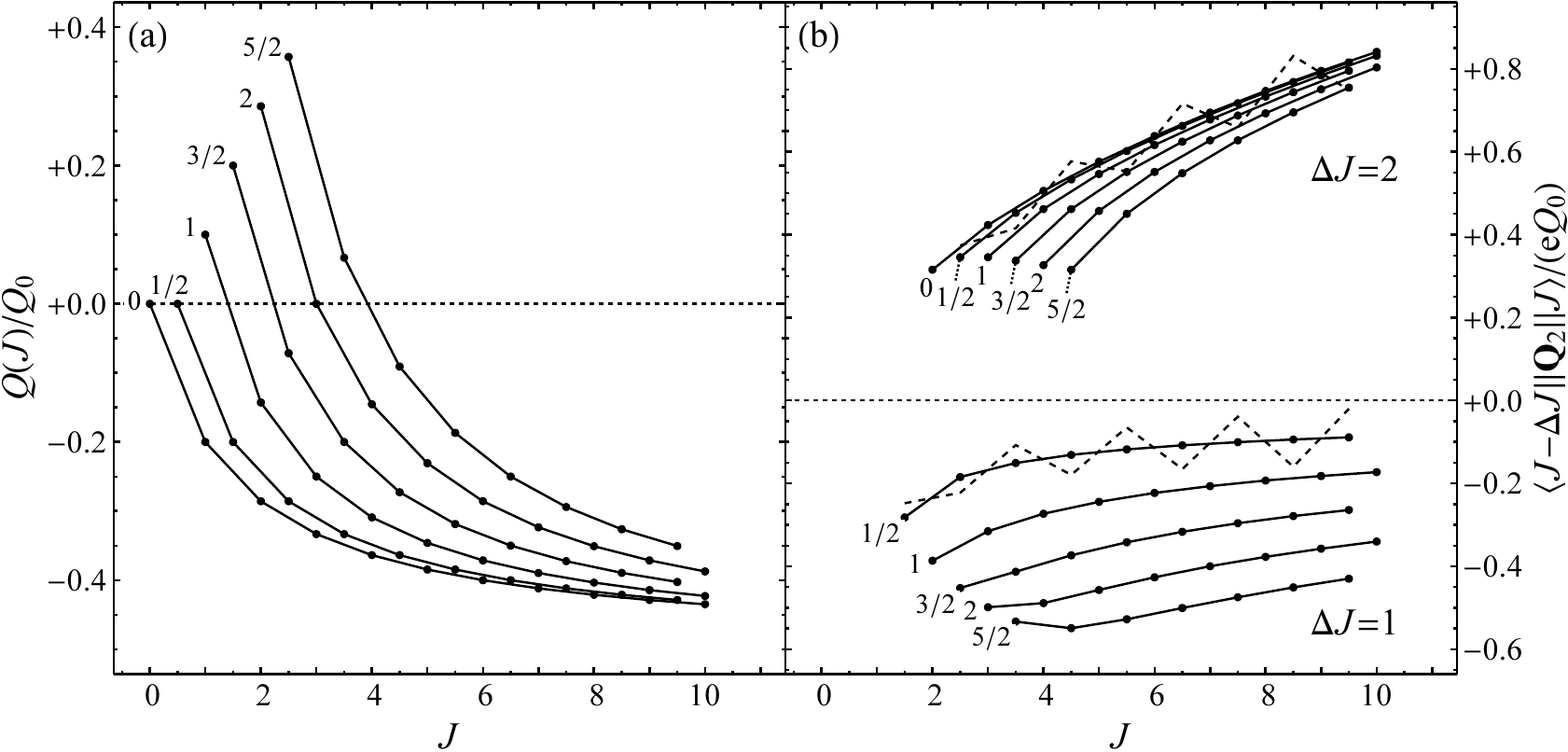}}
\caption{Rotational predictions for electric quadrupole (a)~moments and
(b)~transition reduced matrix elements, within a rotational band,
normalized to the intrinsic quadrupole moment $Q_0$, shown for bands with $0\leq K \leq 5/2$, as indicated.  
The further possibility of staggering of transition strengths within a
$K=1/2$ band~\protect\cite{maris2015:berotor2}
is indicated by
the dotted lines.  Figure from Ref.~\protect\refcite{maris2015:berotor2}.
}
\label{fig-rotor-e2}      
\end{figure}

For the electric quadrupole operator, in particular, the reduced matrix element
between band members (with initial and final angular momenta  $J_i$ and $J_f$, respectively) follows
the relation\footnote{Notations for electromagnetic
observables and the convention for reduced matrix
elements are as defined in Sec.~II of
Ref.~\protect\refcite{maris2015:berotor2}.  Also see 
this reference for discussion of a cross term omitted
from~(\ref{eqn-MEE2}), which may arise for bands with $K=1/2$ or~$1$.}
\begin{equation}
\label{eqn-MEE2}
\trme{\Psi_{J_fK}}{\Qvec_2}{\Psi_{J_iK}}
= 
(2J_i+1)^{1/2}
~
\underbrace{\tcg{J_i}{K}{2}{0}{J_f}{K}}_{\text{Rotational}}
~
\underbrace{\tme{\phi_K}{Q_{2,0}}{\phi_K}}
_{\text{Intrinsic ($\propto eQ_0$)}}.
\end{equation}
The value depends on the particular band members involved, $J_i$ and $J_f$, only
through the Clebsch-Gordan coefficient, while the specific structure of the
intrinsic state enters only through the intrinsic quadrupole moment
$eQ_0\equiv(16\pi/5)^{1/2}\tme{\phi_K}{Q_{2,0}}{\phi_K}$. 

All electric quadrupole
moments $Q(J)$
and reduced transition
probabilities $B(E2;J_i\rightarrow J_f)$
within a given band are therefore uniquely related to each other
via~(\ref{eqn-MEE2}), simply from the assumption of
rotation, with their overall normalization determined by $Q_0$.  That is,
\begin{equation}
\label{eqn-Q}
Q(J)=\frac{3K^2-J(J+1)}{(J+1)(2J+3)}Q_0,
\end{equation}
and
\begin{equation}
\label{eqn-BE2}
B(E2;J_i\rightarrow J_f)=\frac{5}{16\pi} \tcg{J_i}{K}{2}{0}{J_f}{K}^2 (eQ_0)^2.
\end{equation}
Experimental transition strengths are customarily expressed
in terms of the \textit{unsigned} reduced transition
probabilities~--- or $B(E2)$ values~--- as given in~(\ref{eqn-BE2}), since phase information on the
 matrix elements is not normally experimentally
accessible.  However, in the rotational analysis of \textit{ab initio} wave
functions it is more informative to consider the \textit{signed} (unsquared)
reduced matrix elements~(\ref{eqn-MEE2}) directly,  to retain further
meaningful phase information (as illustrated in Sec.~\ref{sec-be-7be}).
The expected rotational relations for electric quadrupole moments and
transition reduced matrix
elements are
summarized graphically in Fig.~\ref{fig-rotor-e2}, for different values of $K$.

The
rotational relation~(\ref{eqn-MEE2}) is equally valid whether we take
the quadrupole operator to be the proton quadrupole tensor
(\textit{i.e.}, the physical electric quadrupole operator) $\Qvec_p$
or the neutron quadrupole tensor $\Qvec_n$.  The matrix elements of
these two operators provide valuable complementary information for
investigating rotation in \textit{ab initio} calculations~--- despite the
comparative (though not complete~\cite{iwasaki2000:12be-pscatt-deformation}) inaccessability of neutron quadrupole observables in
traditional experimental analyses.

\begin{figure}
\centerline{\includegraphics[width=\hsizeforfig]{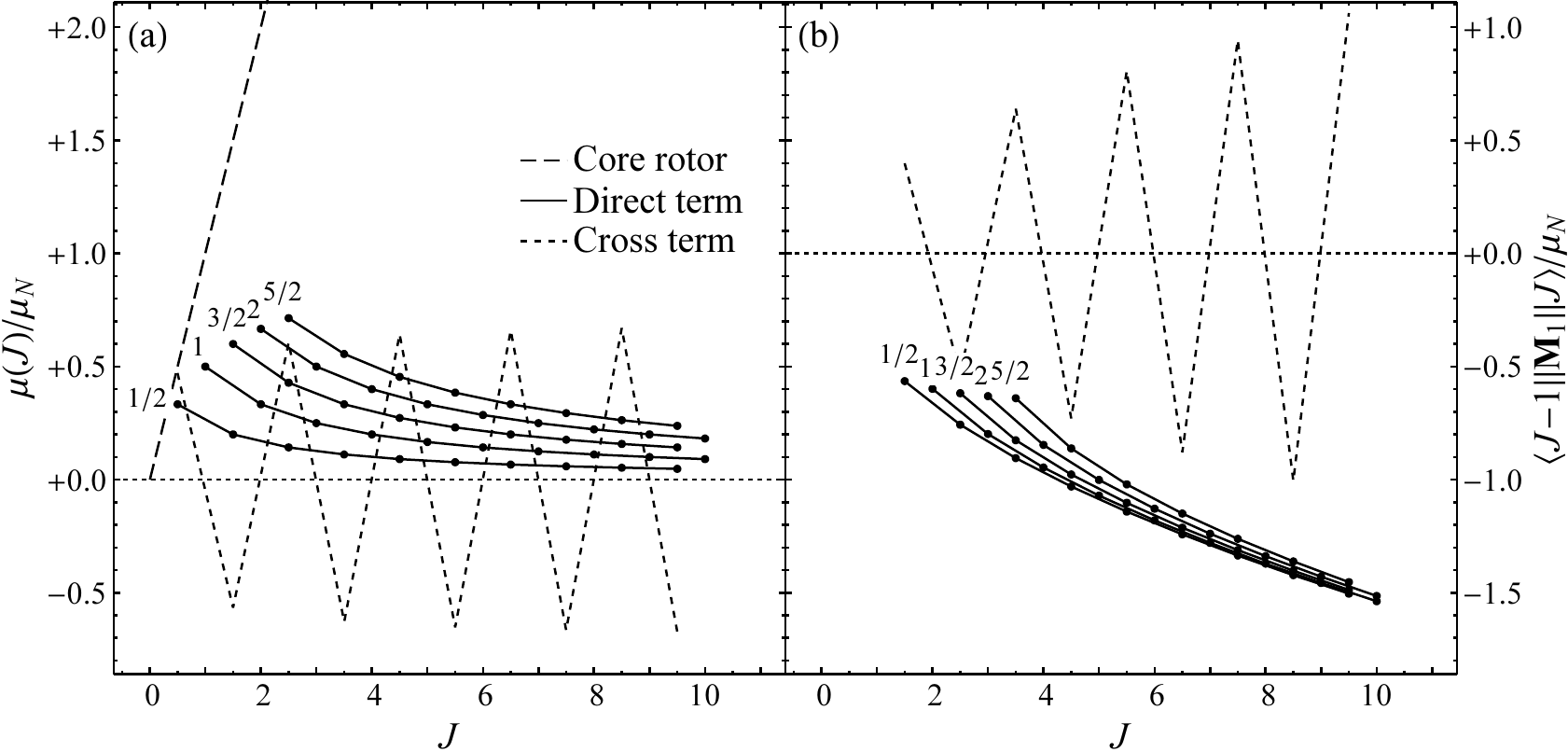}}
\caption{Rotational predictions for each of the terms in~(\ref{eqn-MEM1})
contributing  to magnetic dipole (a)~moments and
(b)~transition reduced matrix elements, within a
rotational band: the core rotor term for dipole
moments (dashed line),
direct term (solid lines, $K\geq1/2$ only, as indicated), and cross term (dotted
lines, $K=1/2$ only).  For purposes of comparison, the curves are
shown with normalizations given by $g_R=1$ and intrinsic matrix elements
equal to $[3/(4\pi)]^{1/2} \mu_N$.  Figure from Ref.~\protect\refcite{maris2015:berotor2}.
}
\label{fig-rotor-m1}      
\end{figure}

Magnetic dipole moments and transitions are deduced from
reduced matrix elements of the magnetic dipole operator.  The
rotational predictions are based on the assumption of separation of
the nucleus into a deformed rotational core, which contributes through an effective
dipole operator simply proportional to
$\Jvec$, plus extra-core nucleons, which contribute through a residual magnetic dipole operator $\Mvec'$.  The
result is a somewhat more complicated rotational expression
\begin{equation}
\label{eqn-MEM1}
\begin{lgathered}
\trme{\psi_{J_fK}}{\Mvec_1}{\psi_{J_iK}}
=
\sqrt{\frac{3}{4\pi}}g_R \mu_N \trme{J_f}{\Jvec}{J_i}
\delta_{J_iJ_f}
\\
\qquad\qquad  
+
({2J_i+1})^{1/2}\Bigl[
\tcg{J_i}{K}{1}{0}{J_f}{K} \tme{\phi_K}{M_{1,0}'}{\phi_K}
 \\
\qquad\qquad \qquad\qquad 
  +\delta_{K,1/2}(-)^{J_i+1/2}
  \tcg{J_i,}{-\tfrac12,}{1,}{1}{J_f}{\tfrac12} \tme{\phi_{1/2}}{M_{1,1}'}{\phi_{\overline{1/2}}}\Bigr],
\end{lgathered}
\end{equation}
for which corresponding simplified expressions for dipole moments
$\mu(J)$ or $\Delta J=1$ transitions may be found in Sec.~II\,D of
Ref.~\protect\refcite{maris2015:berotor2}.
The essential point for purposes of rotational analysis is to note that the
rotational predictions involve three parameters: a core rotational
gyromagnetic ratio $g_R$ (affecting only moments), a direct intrinsic
matrix element $\tme{\phi_K}{M_{1,0}'}{\phi_K}$, and a cross term
intrinsic matrix element
$\tme{\phi_{1/2}}{M_{1,1}'}{\phi_{\overline{1/2}}}$ (for $K=1/2$
bands).   The contributions  to the magnetic
dipole matrix
elements from these various terms are
summarized graphically in Fig.~\ref{fig-rotor-m1}, for different values of $K$.

The physical magnetic dipole operator (excluding meson-exchange currents) is the particular
linear combination
of orbital/spin and proton/neutron angular  momentum operators
\begin{equation}
\label{eqn-m1-D}
\Mvec_1=\sqrt{\frac{3}{4\pi}} \mu_N  \bigl(
\glp \Lvec_p + \gln \Lvec_n + \gsp \Svec_p + \gsn \Svec_n
\bigr)
,
\end{equation}
where the physical gyromagnetic ratios are $\glp=1$,
$\gln=0$, $\gsp\approx 5.586$, and $\gsn \approx-3.826$.  However,
the rotational results~(\ref{eqn-MEM1}) apply to each term
independently (and to any such linear combination).  Therefore, magnetic dipole matrix
elements may be calculated and analyzed considering each of these different \textit{dipole
terms}~\cite{suhonen2007:nucleons-nucleus} individually, to separately probe the
orbital and spin angular momentum structure of
rotation.\footnote{Specifically, the magnetic dipole observables
  quoted for each of these operators will be obtained by setting the
  corresponding gyromagnetic ratio to unity, \textit{i.e.}, using $M1$
  operators defined for each
  dipole term as $\Dlp=[3/(4\pi)]^{1/2}
\mu_N \Lvec_p$,
$\Dln=[3/(4\pi)]^{1/2} \mu_N \Lvec_n$,
\textit{etc.}~\protect\cite{maris2015:berotor2}  Of course, if
one were willing to move further away from traditional notation for
magnetic dipole observables, one could just as well quote
matrix elements of $\Lvec_p$, $\Lvec_n$, \textit{etc.}, directly.}
The magnetic dipole moment or magnetic dipole transition matrix
element pertinent to physical electromagnetic transitions can, of
course, be recovered as the particular linear
combination given in~(\ref{eqn-m1-D}).


\section{Emergence of rotational bands  in the $\isotope{Be}$ isotopes}
\label{sec-be}


\subsection{Rotation in $\isotope[8]{Be}$}
\label{sec-be-8be}

\begin{figure}[tp]
\centerline{\includegraphics[width=\hsizeforfig]{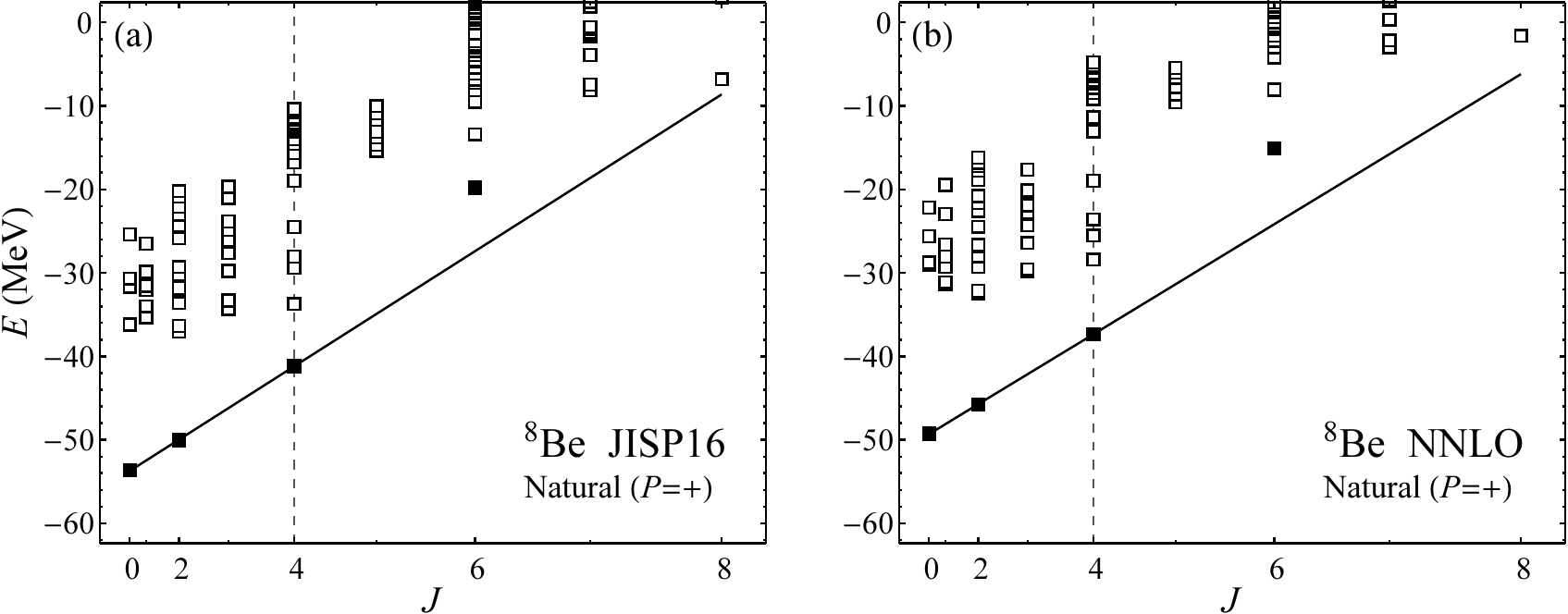}}
\caption{Energy eigenvalues for states in  the
natural parity space of
$\isotope[8]{Be}$, as obtained with the JISP16~(left) and NNLO~(right) nucleon-nucleon interactions.  Energies are plotted with
respect to an angular momentum axis which is scaled to be linear in
$J(J+1)$.
Solid symbols indicate candidate 
band members.  Lines
indicate the corresponding fits for rotational energies~(\ref{eqn-EJ-stagger}).
Vertical dashed lines indicate the maximal angular
momentum accessible within the lowest harmonic oscillator
configuration (or valence space).  From
  calculations with $\Nmax=10$ at $\hw=20\,\MeV$.}
\label{fig-levels-8be0}      
\end{figure}

Let us begin with the simplest case, that of the even-even nucleus
$\isotope[8]{Be}$.  The level energies
obtained in \textit{ab initio} NCCI calculations of $\isotope[8]{Be}$
are shown in Fig.~\ref{fig-levels-8be0}.  

Traditionally, two competing structural descriptions may be
invoked~\cite{inglis1953:p-shell}.  In a cluster description, this
nucleus consists of two $\alpha$ particles, which may undergo rotation
analogous to that of a diatomic molecule, resulting in a $K=0$
positive parity yrast rotational band (with $J=0$, $2$, $4$,
$\ldots$).  However, in a conventional shell-model description,
limited to the $p$-shell valence space, only states with angular
momentum $J\leq4$ can be constructed.

Experimentally, $\isotope[8]{Be}$ is unbound, but the ``ground state''
consists of a narrow $J=0$ resonance, which decays by $2\alpha$
breakup~\cite{npa2004:008-010}.  The next excited levels are $J=2$ and
$4$ resonances.  The energies are approximately
consistent with a rotational pattern, with the experimental
$E(4^+)/E(2^+)\approx 3.75(5)$ lying somewhat above the expected
rotational value of $10/3\approx3.33$.  Since the $2\alpha$
decay mode so completely dominates over electromagnetic decay, ratios
of electromagnetic transition matrix elements among these states are not known
experimentally (only the $4^+\rightarrow2^+$ transition has been
observed~\cite{datar2013:8be-radiative}).

Before further interpreting the results in Fig.~\ref{fig-levels-8be0},
a few comments are in order defining the specifics of the calculations
(applicable also to the calculations discussed in subsequent sections for $\isotope[7,9]{Be}$).
The calculations are obtained using two different
realistic nucleon-nucleon interactions.  The JISP16
interaction~\cite{shirokov2007:nn-jisp16}
[Fig.~\ref{fig-levels-8be0}(a)] is a charge-independent two-body
interaction derived from nucleon-nucleon scattering data and adjusted
via a phase-shift equivalent transformation to describe light nuclei
without explicit three-body interactions.\footnote{The present calculations also
include the Coulomb interaction between protons.  It should be noted
that these calculations are therefore not identical to the JISP16 calculations
previously presented in
  Refs.~\refcite{caprio2013:berotor,maris2015:berotor2,caprio2015:berotor-brasov14},
  in which the Coulomb interaction was omitted,  to ensure exact conservation of isospin.  However, the
  primary effect of the Coulomb interaction is simply to induce a
  shift in the overall binding energies, which is irrelevant to the
  analysis of rotational band observables.}  The
\nnloopt{} interaction~\cite{ekstroem2013:nnlo-opt}
[Fig.~\ref{fig-levels-8be0}(b)] is obtained from chiral effective
field theory at next-to-next-to-leading order (NNLO), with low-energy
constants chosen to reproduce nucleon-nucleon scattering phase shifts.

The many-body
Hamiltonian eigenproblem is then solved using the code
MFDn~\cite{sternberg2008:ncsm-mfdn-sc08,maris2010:ncsm-mfdn-iccs10,aktulga2013:mfdn-scalability},
in a proton-neutron $M$
scheme basis~\cite{whitehead1977:shell-methods}.  States of different
parity are solved for separately.
The states shown in Fig.~\ref{fig-levels-8be0} are in
the natural parity space.\footnote{The parity of the lowest allowed oscillator
configuration, or traditional shell model valence space, may be termed
the \textit{natural parity}, and states of natural parity are more
generally built from oscillator
configurations with even numbers of excitations.  The parity obtained by promoting one
nucleon by one shell may be termed the \textit{unnatural parity}, and
states of unnatural parity are
built from oscillator configurations with odd numbers of excitations.
Thus, for even-mass $p$-shell nuclei, such as $\isotope[8]{Be}$,
natural parity is positive parity, while, for odd-mass $p$-shell nuclei, such as $\isotope[7,9]{Be}$,
natural parity is negative parity.}
The calculations in Fig.~\ref{fig-levels-8be0} are obtained for a specific
choice of basis length parameter ($\hw=20\,\MeV$) and trunction ($\Nmax=10$).  They thus may be thought of as taking
a ``snapshot'' of the spectrum along the path to convergence.  

\begin{figure}[tp]
\centerline{\includegraphics[width=\hsizeforfig]{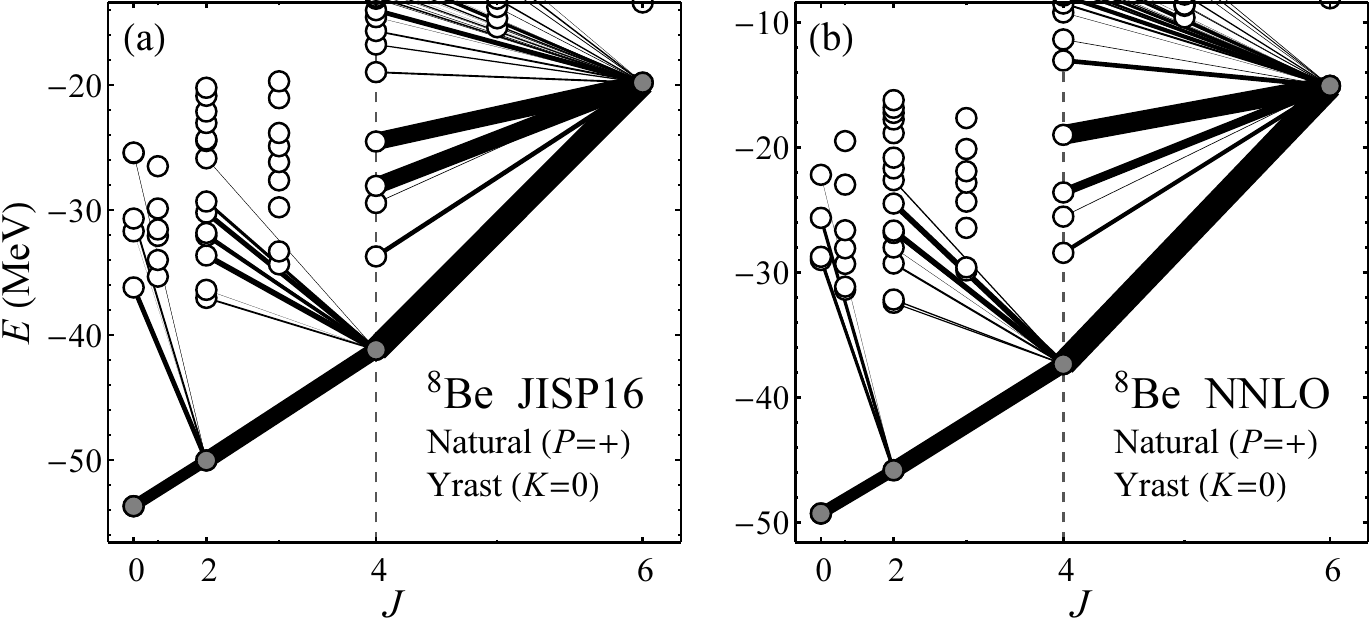}}
\caption{Electric quadrupole transition strengths among levels in the $\isotope[8]{Be}$ natural parity
  space, orignating from yrast band members, as obtained with the
  JISP16~(left) and NNLO~(right) nucleon-nucleon interactions.
  Shaded symbols indicate the initial levels being considered.  All
  angular momentum \textit{decreasing} transitions from the selected
  levels are shown.  Line thicknesses are proportional to the
  magnitude of the reduced matrix element for the transition (also
  conveyed through a gray scale).  From
  calculations with $\Nmax=10$ at $\hw=20\,\MeV$, using the proton
  quadrupole tensor.}
\label{fig-network-8be0}      
\end{figure}

Energies following a rotational pattern are most easily recognized if
plotted against an angular momentum axis which is scaled as $J(J+1)$,
as in Fig.~\ref{fig-levels-8be0},
so that energies in an ideal rotational band lie on a straight line
(or staggered about a straight line, for $K=1/2$).  Rotational bands are most readily identifiable near
the yrast line, where the density of states remains comparatively low.
The band members are identified~--- both in the present discussion of
$\isotope[8]{Be}$ and in the subsequent discussions of other
isotopes~--- on the basis not only of their energies, but also on the
basis of collective enhancement of electric quadrupole transition
strengths among band members.  The strengths of the various electric
quadrupole transitions originating from the candidate band members in
$\isotope[8]{Be}$ are shown in Fig.~\ref{fig-network-8be0}.  

The candidate yrast rotational band members for $\isotope[8]{Be}$ are
indicated (solid symbols) in Fig.~\ref{fig-levels-8be0}, through
$J=6$.  Qualitatively, the situation is similar in the calculations
with either the JISP16 interaction [Fig.~\ref{fig-levels-8be0}(a)] or
NNLO interaction [Fig.~\ref{fig-levels-8be0}(b)].   The yrast $J=0$,
$2$, and $4$ states are well-isolated in energy from the off-yrast
states and lie approximately on a straight line plotted with respect
to $J(J+1)$ (the line shown in Fig.~\ref{fig-levels-8be0} is the best rotational energy fit to the
calculated band members, specifically, with $J\leq4$).  Their energies approximately match the
rotational expectation, with $E(4^+)/E(2^+)\approx3.42$ or $3.46$,
respectively, for the two calculations shown.  While the energy
of the calculated yrast $J=6$ state lies well above the rotational
line (Fig.~\ref{fig-levels-8be0}), the quadrupole transition strengths
(Fig.~\ref{fig-network-8be0}) nonetheless suggest that this state is a member of the yrast band: it
is connected to the yrast $4^+$ band member with collective strength, as well as, less strongly, to other off-yrast $4^+$ states.  Candidate $8^+$ band members
may be identified, as well, but they lie off the yrast line (and outside
the energy range shown in Fig.~\ref{fig-levels-8be0}).

\begin{figure}[tp]
\centerline{\includegraphics[width=\hsizeforfig]{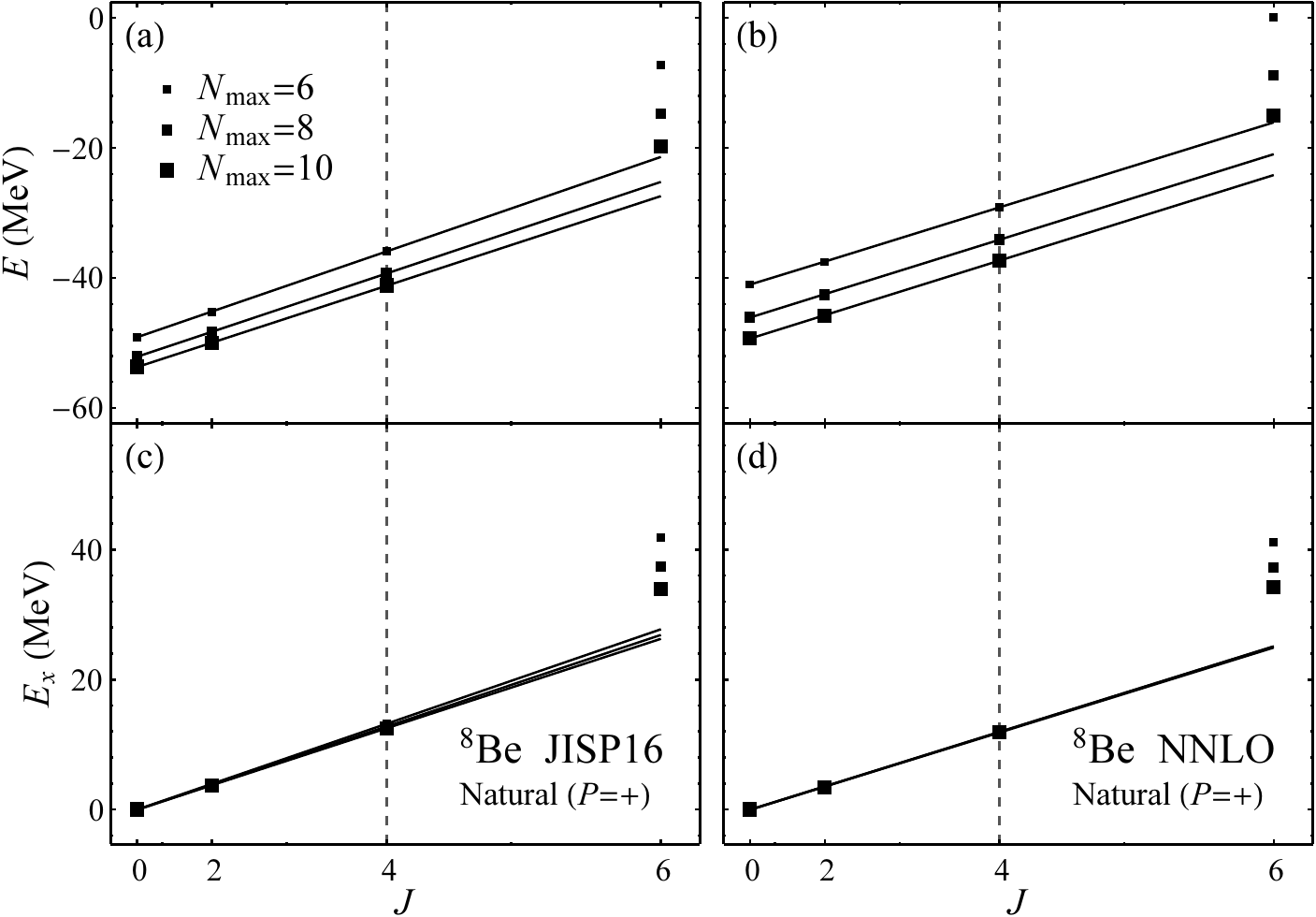}}
\caption{Dependence of calculated energies for
$\isotope[8]{Be}$ natural parity yrast band members on $\Nmax$:
  energy eigenvalues $E$~(top) and excitation energies $E_x$~(bottom),
  as obtained with the JISP16~(left) and NNLO~(right)
  nucleon-nucleon interactions.  
Larger symbols indicate higher $\Nmax$ values.  Lines
indicate the corresponding fits for rotational energies~(\ref{eqn-EJ-stagger}).
Vertical dashed lines indicate the maximal valence angular momentum.  From
  calculations with $\Nmax=6$, $8$, and $10$ at $\hw=20\,\MeV$.}
\label{fig-energy-Nmax-8be0}      
\end{figure}

Let us return to the challenge of convergence (Sec.~\ref{sec-ncci}),
but now with rotational energy patterns in mind.  The underlying
question is how, when the calculated energies are still shifting on an
$\MeV$ scale with increasing basis size, rotational patterns can
nonetheless be reproduced at an $\MeV$ or sub-$\MeV$ scale.  The
$\Nmax$ dependence of the energy eigenvalues is shown for the members
of the yrast band in Fig.~\ref{fig-energy-Nmax-8be0}~(top).  For each
step in $\Nmax$, the calculated energies shift lower by several
$\MeV$, much as already seen for the ground state energy in
Fig.~\ref{fig-conv}(b).  However, it may also be noticed that
the energies of different band members move downward in approximate
synchrony as $\Nmax$ increases (at least for the $J=0$, $2$, and $4$
band members). Thus, the \textit{relative} energies \textit{within}
the band remain comparatively unchanged, as is seen more directly when
we consider excitation energies, in
Fig.~\ref{fig-energy-Nmax-8be0}~(bottom).  Thus, a rotational pattern
in the relative energies remains robustly present, even as the energy
eigenvalues themselves change.  Moreover, the slope or, equivalently,
rotational constant $A$ would seem to be essentially converged.

There is, however, a clear difference in the convergence properties of
the $J=6$ band member, which lies above the maximal angular momentum
($J=4$) accessible within the valence space or, equivalently, maximal
angular momentum accessible within
$\Nmax=0$ NCCI calculations (dashed vertical lines in
Fig.~\ref{fig-energy-Nmax-8be0}).  While the energy of the $J=6$ band
member lies above the rotational expectation, this energy is also
converging downward relative to the energies of the lower band
members.

\begin{figure}[tp]
\centerline{\includegraphics[width=\hsizeforfig]{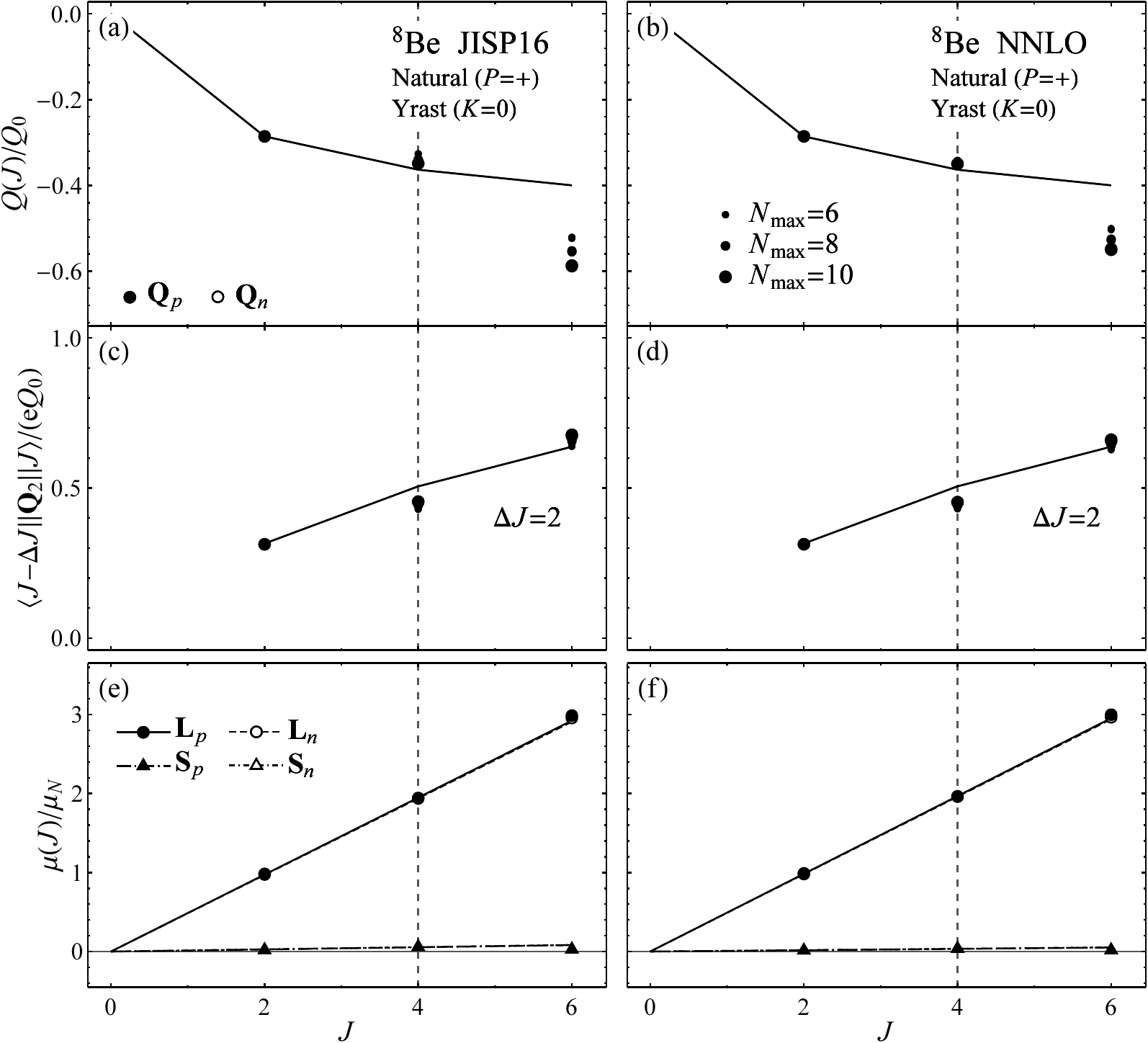}}
\caption{Quadrupole and dipole matrix element observables for the
  $\isotope[8]{Be}$ natural parity yrast band and their dependence on
  the $\Nmax$ truncation: quadrupole moments~(top), quadrupole
  transition reduced matrix elements~(middle), and dipole
  moments~(bottom), as obtained with the JISP16~(left) and
  NNLO~(right) nucleon-nucleon interactions.  Larger symbols indicate higher $\Nmax$ values.  Quadrupole
  observables are normalized to $Q_0$ (see text).  Quadrupole observables
  calculated using both proton and neutron operators and dipole
  observables calculated using all four dipole terms are shown (but
  proton and neutron values nearly coincide in all cases).  The curves indicate
  rotational values (for dipole observables, based on a best fit at highest $\Nmax$). The vertical dashed lines indicate the maximal valence angular
  momentum.  From calculations with $\Nmax=6$,
  $8$, and $10$ at $\hw=20\,\MeV$.}
\label{fig-trans-8be0}      
\end{figure}

Moving now to electromagnetic observables,
recall that the calculated electric quadrupole matrix elements
for $\isotope[8]{Be}$ are far from converged, as we have seen in particular for the transition between the $J=2$ and $J=0$
band members [Fig.~\ref{fig-conv}(f)].  However, rotational
structure is reflected not in the values of these
observables on an \textit{absolute} scale, but rather on the
\textit{ratios} of matrix elements within a band, as dictated by the
Clebsch-Gordan factor in the rotational formula~(\ref{eqn-MEE2}).
(The overall normalization of these matrix elements is then determined
by the intrinsic structure, via the intrinsic quadrupole moment
$Q_0$.)  Let us therefore consider the \textit{relative} values of the
quadrupole moments (proportional to diagonal matrix elements of the
quadrupole operator) within the yrast band of $\isotope[8]{Be}$, in
Fig.~\ref{fig-trans-8be0}~(top) and, similarly, the quadrupole
transition reduced matrix elements (or off-diagonal matrix elements of the
quadrupole operator), in Fig.~\ref{fig-trans-8be0}~(middle).  The
overall normalization $Q_0$ is eliminated by normalizing to one of
these values.  We choose to normalize to the first
nonvanishing quadrupole moment within the band, \textit{i.e.}, of the $J=2$
band member.  That is, the value of $Q_0$ used for normalization
in Fig.~\ref{fig-trans-8be0} is determined from the
  calculated $Q(2)$ via~(\ref{eqn-Q}).  Results are shown both for the JISP16
interaction (at left) and the NNLO interaction (at right).  In each
case, the \textit{relative} quadrupole matrix elements within
the band are seen to be largely converged with respect to $\Nmax$.

The expected rotational values for the quadrupole moments and
transition matrix elements, from~(\ref{eqn-MEE2}) and~(\ref{eqn-Q}), are
given by the curves in Fig.~\ref{fig-trans-8be0}~(top, middle).
Notice, comparing the left and right columns of
Fig.~\ref{fig-trans-8be0}, that the calculated quadrupole observables
obtained with the JISP16 and NNLO interactions are virtually
identical, not only in their resemblance to rotational predictions but
also in the nature of their deviations from the rotational
predictions.  It should be emphasized that the same $Q_0$ values are
used for normalization of the transitions
[Fig.~\ref{fig-trans-8be0}(c) or (d)] as for the quadrupole moments in
the same calculation [Fig.~\ref{fig-trans-8be0}(a) or (b),
  respectively].  Therefore, no free normalization parameter remains
for the transition matrix elements.  For example, since the value of
$Q_0$ used for normalization in Fig.~\ref{fig-trans-8be0} is
determined from $Q(2)$, the proximity of the lowest calculated
transition data point to the rotational curve indicates that the
\textit{ab initio} calculations exhibit an agreement between $Q(2)$
and $B(E2;2^+\rightarrow0^+)$ consistent with adiabatic rotation.
There is a break from the rotational predictions in the quadrupole
moments at $J=6$~--- the quadrupole moment $Q(6)$ has the expected
sign but is nearly half again as large in magnitude as expected from
the rotational formula (and still increasing in magnitude with
increasing $\Nmax$).  On the other hand, the transition matrix
element from this $J=6$ band member is still reasonably consistent with
the rotational formula.

Since the difference between proton and neutron quadrupole observables
will take on more significance going forward to the other
$\isotope{Be}$ isotopes (Secs.~\ref{sec-be-7be}--\ref{sec-be-9be}), it
is worth noting that, for $\isotope[8]{Be}$, the matrix elements
calculated using the proton and neutron quadrupole operators~---
$\Qvec_p$ and $\Qvec_n$ (Sec.~\ref{sec-rot})~--- are almost identical,
as a result of the approximate proton-neutron symmetry of the system.
Quadrupole moments and matrix elements calculated using the proton
(solid symbols) and neutron (open symbols) quadrupole tensors are, in
principle, both shown in Fig.~\ref{fig-trans-8be0}, but the data
points are almost entirely indistinguishable on the plot.  The proton
and neutron quadrupole observables are normalized separately in
Fig.~\ref{fig-trans-8be0}, \textit{i.e.}, the proton and neutron
intrinsic quadrupole moments ($Q_{0,p}$ and $Q_{0,n}$, respectively)
are determined independently.  Most of the difference in the
calculated proton and neutron observables is embodied in this
normalization, through an $\sim1\%$ difference in $Q_{0,p}$ and $Q_{0,n}$.

There are no magnetic dipole transitions to consider within a $K=0$ band,
since the angular momenta of successive band members differ by $2$.
However, we may still examine the dipole moments for the rotational
band members in $\isotope[8]{Be}$, as shown in
Fig.~\ref{fig-trans-8be0}~(bottom).  As a particular special case
of~(\ref{eqn-MEM1}), these are expected to vary linearly with $J$, as
$\mu(J)=g_R\mu_N J$.  The dipole moments calculated with the orbital
angular momentum dipole terms [circles in
  Fig.~\ref{fig-trans-8be0}~(bottom)] do indeed closely follow such a
linear relation.  The values are well-converged, with a slope
$g_R\approx0.49$.\footnote{The traditional collective result for the
  gyromagnetic
  ratio~\protect\cite{kurath1961:p-shell-magnetic-moments} is
  $g_R=0.5$, obtained if we assume identical contributions
  $\Lvec_p=\Lvec_n$ from the proton and neutron orbital angular
  momenta and no contribution from spin angular momenta.}
Note that the calculated dipole moments agree with the simple linear
rotational formula even for the $J=6$ band member, for which the
calculated energy and quadrupole moment were not as clearly consistent
with a rotational picture.

For the spin dipole terms [triangles in
  Fig.~\ref{fig-trans-8be0}~(bottom)], the
moments are nearly vanishing.  
The dipole moments calculated with the
different orbital/spin and proton/neutron dipole terms may be
interpreted as ``spin contributions'' coming from these different
operators~\cite{maris2013:ncsm-pshell}, since they measure the
projection of the orbital or spin angular momentum onto the total
angular momentum.  The near-vanishing contribution from the
intrinsic spins is consistent with an $\alpha$-clustering picture,
where the spins pair to zero angular momentum.

Returning to the complementary shell model description, it may be
noted that Elliot $\grpsu{3}$
symmetry~\cite{elliott1958:su3-part1,harvey1968:su3-shell} and the LS
coupling scheme play significant organizing roles in the structure of
$p$-shell nuclei~\cite{millener2001:light-nuclei}.  \textit{Ab
  initio} calculations of the $\isotope[8]{Be}$ ground state  in an $\grpsu{3}$ coupling
scheme have been reported in Ref.~\refcite{dytrych2013:su3ncsm}, using a next-to-next-to-next-to-leading-order (\nthreelo{}) chiral
interaction~\cite{entem2003:chiral-nn-potl}.  It is found that the dominant contribution to the ground
state wave function arises from the $(4,0)$ irreducible
representation (irrep) of  $\grpsu{3}$, paired with intrinsic spin
contributions which all vanish, \textit{i.e.}, $(S_p,S_n,S)=(0,0,0)$.  The $(4,0)$ irrep
of $\grpsu{3}$ contains angular momentum states with $L=0$, $2$, and
$4$, corresponding to a truncated $K=0$ rotational band.

To briefly summarize these observations, from the \textit{ab initio}
calculations there are clear and consistent indications of rotation,
in the simplest example of $\isotope[8]{Be}$, based on patterns in
energies and electromagnetic observables.  The 
$K=0$ ground state band is qualitatively consistent with an
$\alpha$-$\alpha$ structure, but discontinuities in
observables at the maximal valence angular momentum suggest that the
spherical shell structure (and shell model $p$-shell description) may retain physical relevance.


\subsection{Rotation in $\isotope[7]{Be}$}
\label{sec-be-7be}

The most distinctive and well-developed
rotational band structures are observed in calculations for odd-mass
nuclei. Given the same range of excitation energies and angular
momenta, the low-lying $\Delta J=1$ bands in the odd-mass nuclei
provide a richer set of energy and electromagnetic observables.  
Yrast and near-yrast states
yield the most immediately recognizable sets of candidate
band members, so our focus will be on these states.
The rotational bands in $\isotope[7]{Be}$ (this section) and
$\isotope[9]{Be}$ (Sec.~\ref{sec-be-9be})
serve as illustrative cases.
Experimental
counterparts  for these
calculated rotational bands may be identified~\cite{npa2002:005-007,npa2004:008-010,bohlen2008:be-band,maris2015:berotor2}.
\begin{figure}[tp]
\centerline{\includegraphics[width=\hsizeforfig]{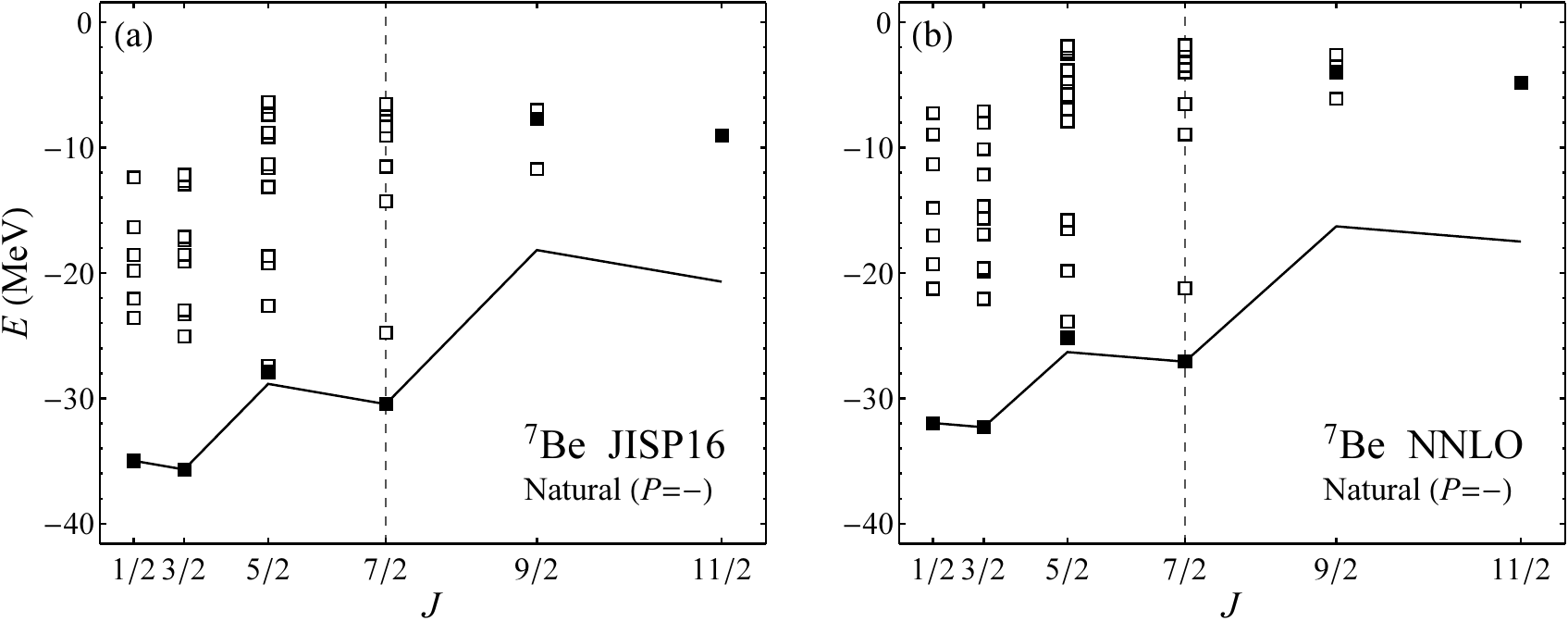}}
\caption{Energy eigenvalues for states in  the
natural parity space of
$\isotope[7]{Be}$, as obtained with the JISP16~(left) and NNLO~(right)
nucleon-nucleon interactions.  
See Fig.~\ref{fig-levels-8be0} caption for discussion of plot
contents and labeling.   From
  calculations with $\Nmax=10$ at $\hw=20\,\MeV$.}
\label{fig-levels-7be0}      
\end{figure}
\begin{figure}[tp]
\centerline{\includegraphics[width=\hsizeforfig]{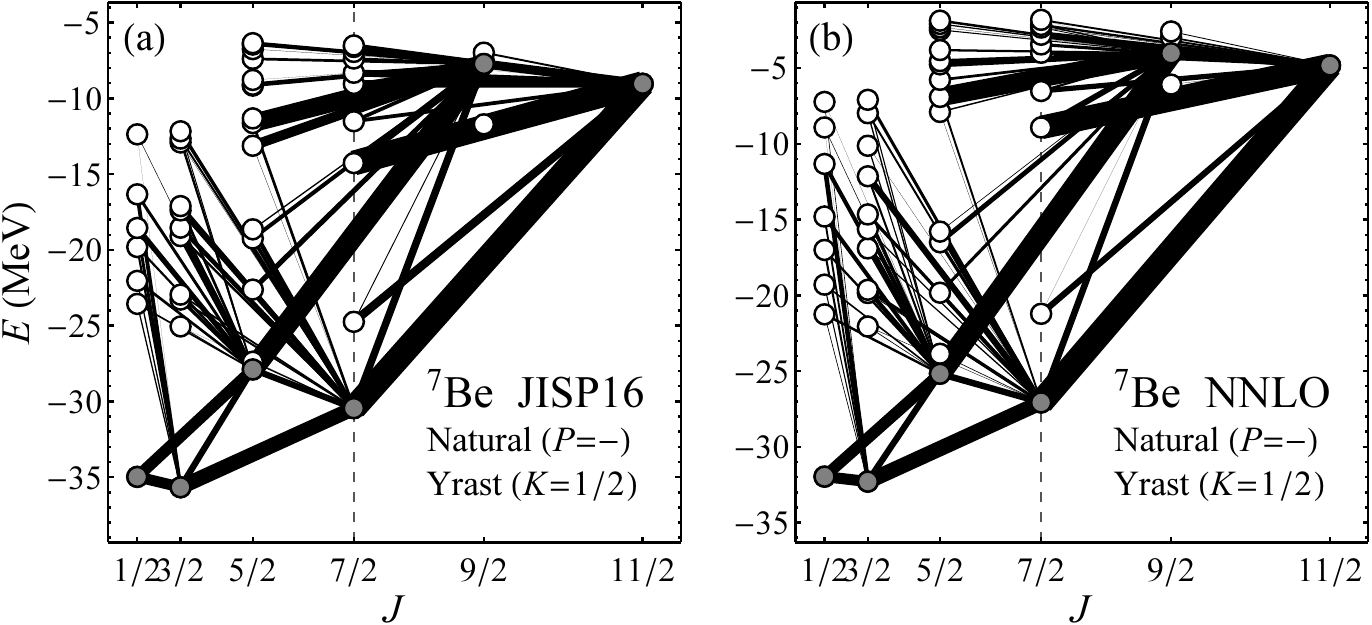}}
\caption{Electric quadrupole transition strengths among levels in the $\isotope[7]{Be}$ natural parity
  space, orignating from yrast band members, as obtained with the
  JISP16~(left) and NNLO~(right) nucleon-nucleon interactions.
See Fig.~\ref{fig-network-8be0} caption for discussion of plot
contents and labeling.  From
  calculations with $\Nmax=10$ at $\hw=20\,\MeV$, using the proton
  quadrupole tensor.}
\label{fig-network-7be0}      
\end{figure}

The low-lying levels calculated in $\isotope[7]{Be}$ are shown in
Fig.~\ref{fig-levels-7be0}.  A $K=1/2$ yrast band is identified, again
through a combination of rotational energies and collectively enhanced
transition strengths.  The quadrupole transitions, shown in
Fig.~\ref{fig-network-7be0}, obey the characteristic pattern for a
$K=1/2$ band implied by the rotational model
[Fig.~\ref{fig-rotor-e2}(b)]: stronger $\Delta J=2$ transitions and
comparatively weak (though still collective) $\Delta J=1$ transitions.
The energy staggering is such that the $J=1/2$, $5/2$, $\ldots$ levels
are raised in energy, and the $J=3/2$, $7/2$, $\ldots$ levels are
lowered (this direction for the staggering corresponds to a negative value
of the Coriolis decoupling parameter $a$).  Note that the staggering
is sufficiently pronounced that the two lowest-$J$ band members are
inverted, as is experimentally observed.

Comparing the calculated energies with the rotational
formula~(\ref{eqn-EJ-stagger}), it may be seen that the energies of
the band members through the highest angular
momentum accessible in the valence space ($J=7/2$) are reasonably consistent
with the rotational formula.  (The line
in Fig.~\ref{fig-levels-7be0} represents the predictions of the
rotational formula, with band energy parameters $E_0$, $A$, and $a$
extracted from the energies of the three lowest-energy band members,
\textit{i.e.}, $J=1/2$, $3/2$, and $7/2$.)  Although a second $J=5/2$
state lies within $\sim1\,\MeV$ of the yrast $J=5/2$ state, in both
calculations, the lack of enhanced transitions
(Fig.~\ref{fig-network-7be0}) suggests negligible mixing of this
``spectator'' state with the yrast band member.  

At higher angular momenta, $J=9/2$ and $11/2$ states are calculated to
have collective quadrupole transitions (Fig.~\ref{fig-network-7be0})
to the lower band members, suggesting their inclusion as band members.
The energy staggering is such that the $J=9/2$ band member lies off
the yrast line.  (Intriguingly, these states also have enhanced
transitions to certain other, high-lying $J=5/2$ and $7/2$ states, at
excitation energies which seem to be roughly consistent between the
JISP16 and NNLO{} calculations.)  While the energies of these states
lie above the rotational expectation (Fig.~\ref{fig-levels-7be0}),
these energies are also converging downward more rapidly than those of the
lower band members, much as seen above for the $J=6$ band member in
$\isotope[8]{Be}$ (Sec.~\ref{sec-be-8be}).

\begin{figure}[tp]
\centerline{\includegraphics[width=\hsizeforfig]{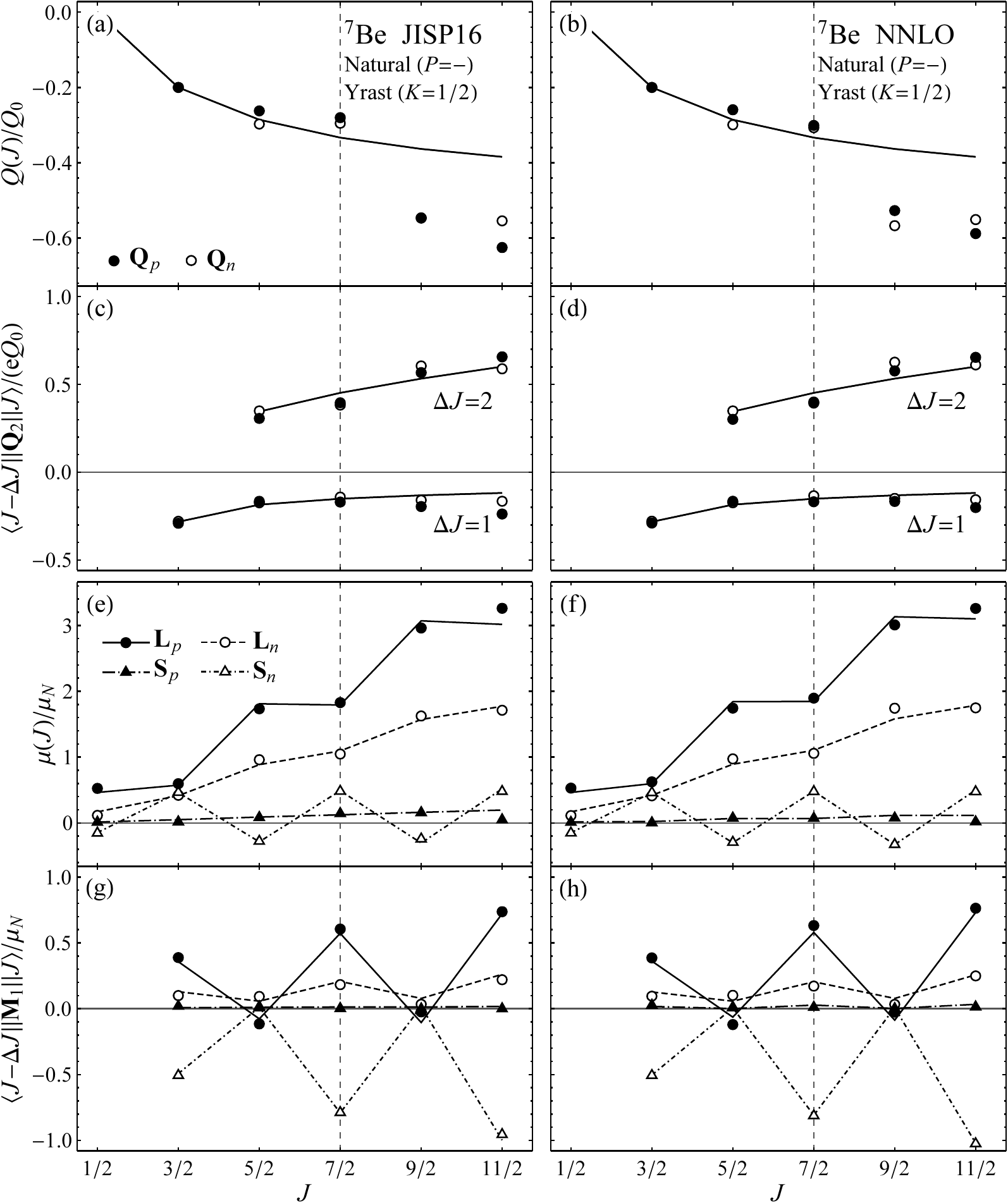}}
\caption{Quadrupole and dipole matrix element observables for the
  $\isotope[7]{Be}$ natural parity yrast band: quadrupole
  moments~(first row), quadrupole transition reduced matrix
  elements~(second row), dipole moments~(third row), and dipole
  transition reduced matrix elements~(fourth row),
as obtained with the
  JISP16~(left) and NNLO~(right) nucleon-nucleon interactions.
See Fig.~\ref{fig-trans-8be0} caption for discussion of plot
contents and labeling.  From
  calculations with $\Nmax=10$ at $\hw=20\,\MeV$.
}
\label{fig-trans-7be0}      
\end{figure}

A detailed test of the rotational description for these candidate
$\isotope[7]{Be}$ band members is obtained by comparing the many
electric quadrupole moments and transition matrix elements among these
states with the rotational expectation from~(\ref{eqn-MEE2})
and~(\ref{eqn-Q}), as shown in the upper two rows of
Fig.~\ref{fig-trans-7be0}.  Recall that only a single normalization
constant, the intrinsic quadrupole moment $Q_0$, enters into the
rotational predictions for the electric quadrupole moments and
transitions, which are shown normalized to $Q_0$ in
Fig.~\ref{fig-trans-7be0}.  (We have fixed $Q_0$ based on the lowest
nonvanishing quadrupole moment, that of the $J=3/2$ state.)  Beyond
this choice of normalization, agreement or disagreement of the
\textit{ab initio} calculated values with the rotational curves is
entirely a test of the rotational picture.  Quadrupole moments in
Fig.~\ref{fig-trans-7be0} are calculated using both the proton (solid
symbols) and neutron (open symbols) quadrupole tensors
(Sec.~\ref{sec-rot}).  The proton and neutron quadrupole moments are
normalized separately, since no \textit{a priori} relation exists
between the intrinsic matrix elements of the $\Qvec_p$ and $\Qvec_n$
operators.  (In some cases, data points for the neutron and proton
results may not be separately visible in these figures, when the
values are so close as to be indistinguishable.)

We may observe an essentially similar behavior to that noted earlier
for quadrupole observables in the $\isotope[8]{Be}$ yrast band
(Sec.~\ref{sec-be-8be}), though now with the added richness of $\Delta
J=1$ transitions.  The quadrupole moments
[Fig.~\ref{fig-trans-7be0}~(first row)] are consistent with a
rotational picture up to the maximal valence angular momentum $J=7/2$.
There is again a modest discontinuity in the quadrupole moments
(increasing by about half again over the rotational expectation) above
this angular momentum.  Although the $\Nmax$ dependence is not shown
in Fig.~\ref{fig-trans-7be0}, the values of these quadrupole moments
(even relative to those of the rest of the band, \textit{i.e.},
normalized to $Q_0$) are poorly converged with $\Nmax$.  The
quadrupole transition matrix elements
[Fig.~\ref{fig-trans-7be0}~(second row)] remain largely consistent
with the rotational expectations throughout the candidate band, up to
$J=11/2$.

Likewise, note the remarkable level of consistency between the JISP16
[Fig.~\ref{fig-trans-7be0}~(left)] and NNLO
[Fig.~\ref{fig-trans-7be0}~(right)] calculations of these observables.
The similarity of these calculations lies not just in their mutual
overall agreement with the rotational predictions, but in the nature
of their deviations from the rotational formula and the
sense of the splittings between the values of proton and neutron
matrix elements (excepting, perhaps,
certain details for the highest-$J$ band members).

The calculations in Fig.~\ref{fig-trans-7be0} test not just the
magnitudes of the moments and matrix elements, but also their signs.
There are some ambiguities in the rotational predictions for the signs
of the reduced matrix elements, due to the arbitrary phases entering
into the definition of each eigenstate of a Hamiltonian operator (or,
equivalently, the arbitrary overall sign arising on each eigenvector
obtained in the numerical diagonalization of a Hamiltonian matrix).
However, even once phase ambiguities are taken into account, a rich
set of predicted correlations between signs of matrix elements for
electric quadrupole ($\Delta J=1$ and $\Delta J=2$) and magnetic
dipole transitions remains, as detailed in Sec.~III\,C of
Ref.~\protect\refcite{maris2015:berotor2}.  Since moments are deduced
from \textit{diagonal} matrix elements of the transition operator,
they are invariant under the arbitrary sign choices in the definitions
of eigenstates.  The sign of a transition matrix element varies with
the sign choices on the initial and final states, $\sigma_J$ and
$\sigma_{J'}$, respectively, as the product $\sigma_{J'} \sigma_J$.

Matching a small subset of the \textit{ab initio} calculated
transition matrix elements (say, the $\Delta J=1$ proton quadrupole
matrix elements) to the signs conventionally adopted in the rotational
description (Fig.~\ref{fig-rotor-e2}) suffices to completely fix 
arbitrary signs.  It is then meaningful to compare the signs of all
other transition matrix elements~--- the proton and neutron
quadrupole matrix elements, and all magnetic dipole terms~--- with the
rotational predictions.  The signs obtained in the \textit{ab initio}
calculations of Fig.~\ref{fig-trans-7be0} are uniformly consistent
with the rotational picture.

We turn now to comparing the many magnetic dipole moments and transition
matrix elements among these states with the rotational expectation
from~(\ref{eqn-MEM1}), as shown in the lower two rows of
Fig.~\ref{fig-trans-7be0}.  The four distinct ``dipole moments''
calculated for each band member in $\isotope[7]{Be}$, obtained with
the four different dipole terms, are
shown in Fig.~\ref{fig-trans-7be0}~(third row), while the four distinct sets
of $\Delta J=1$ dipole transition matrix elements are shown in
Fig.~\ref{fig-trans-7be0}~(fourth row). The lines indicate the
rotational predictions from~(\ref{eqn-MEM1}), with parameters
determined to provide a best fit to
the calculated moments and transitions (specifically, considering 
the band members with $J\leq7/2$).
These parameters are determined
independently, for each dipole term operator (and, of course, for the
calculations with different interactions).

Recall that there are only three parameters in the rotational
predictions~(\ref{eqn-MEM1}).  The core gyromagnetic ratio $g_R$ is
responsible for the overall linear trend in the dipole moments, which
is the dominant contribution for the orbital dipole terms [circles in
Fig.~\ref{fig-trans-7be0}~(third row)]. Then, for a $K=1/2$ band, there are two
relevant intrinsic matrix elements, where the second of these, or cross term
in~(\ref{eqn-MEM1}), contributes the staggering of the
values as a function of $J$.  
Thus, there are only three parameters but many more (eleven)
calculated values (for each dipole term and for each calculation in
Fig.~\ref{fig-trans-7be0}), and the \textit{ab initio}
calculated values appear to be highly consistent with a rotational
pattern.  Moreover, the senses of the deviations which do arise appear to be
remarkably consistent between the JISP16 and NNLO calculations.
Notice the  near-vanishing proton spin contributions~--- these would
be consistent, for instance, with a
structure in which proton spins are paired to yield zero total proton
spin angular momentum.

The rotational formula for magnetic dipole moments and transition
matrix elements provided by~(\ref{eqn-MEM1}) is the result of the
classic rotational interpretation, formulated for heavier nuclei.  As
noted in Sec.~\ref{sec-rot}, the basic assumption is that the nucleus
separates into a deformed rotating core and residual extra-core
particles.  However, consistency with a model does not imply that the
model provides the sole successful description of the physical system,
nor that the model uniquely provides the correct underlying physical
interpretation.  For $\isotope[7]{Be}$, in particular, it might be
natural, in a cluster description, to consider the nucleus either as
an $\alpha$-$\isotope[3]{He}$ dimer or as $\isotope[8]{Be}$ coupled to
a neutron hole~\cite{inglis1953:p-shell}.


\subsection{Rotation in $\isotope[9]{Be}$}
\label{sec-be-9be}

\begin{figure}[tp]
\centerline{\includegraphics[width=\hsizeforfig]{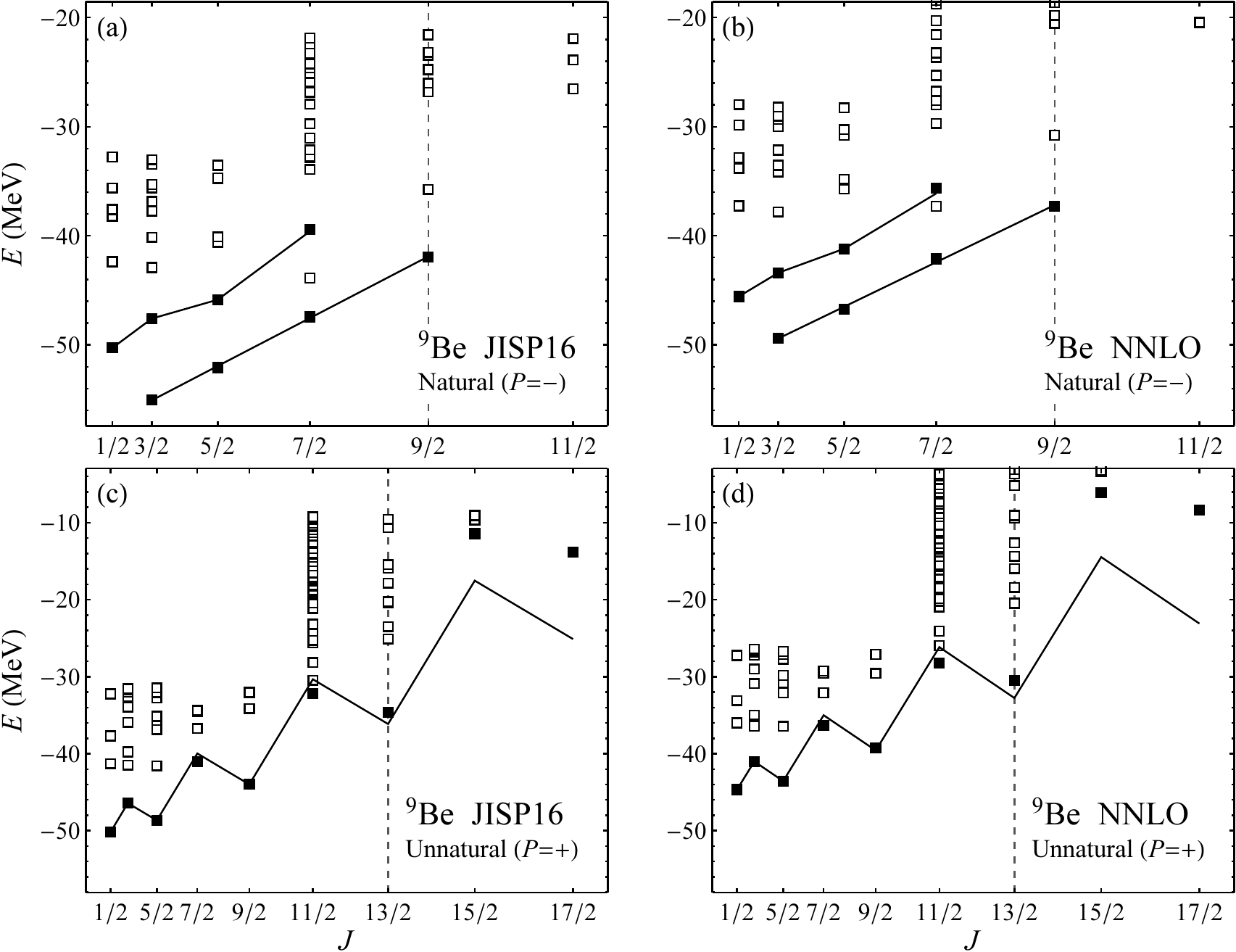}}
\caption{Energy eigenvalues for states in  the
natural parity space of
$\isotope[9]{Be}$ (top) and unnatural parity space of
$\isotope[9]{Be}$ (bottom), as obtained with the JISP16~(left) and NNLO~(right)
nucleon-nucleon interactions.  
See Fig.~\ref{fig-levels-8be0} caption for discussion of plot
contents and labeling.   
From
  calculations with $\Nmax=10$ (for natural parity) and $\Nmax=11$
  (for unnatural parity) at $\hw=20\,\MeV$.}
\label{fig-levels-9be01}      
\end{figure}

The isotope $\isotope[9]{Be}$ has a natural interpretation in a
cluster picture, as consisting of $\isotope[8]{Be}$ plus a neutron,
that is, as an $\alpha$-$\alpha$ dimer with a covalent neutron shared
between the $\alpha$ clusters~\cite{inglis1953:p-shell}.  Although the ground state is
stable, all excited states are resonances, lying
above the $\alpha+\alpha+n$ decay threshold~\cite{npa2004:008-010}.
The unnatural (positive) parity states begin at low excitation energy
relative to the natural (negative) parity states: the ground state is
$3/2^-$, but the first excited state, at an excitation energy of under $2\,\MeV$, is $1/2^+$.
We therefore show, in Fig.~\ref{fig-levels-9be01}, the energies of the
eigenstates obtained in both the natural
[Fig.~\ref{fig-levels-9be01}~(top)] and unnatural
[Fig.~\ref{fig-levels-9be01}~(bottom)] parity spaces.

\begin{figure}[tp]
\centerline{\includegraphics[width=\hsizeforfig]{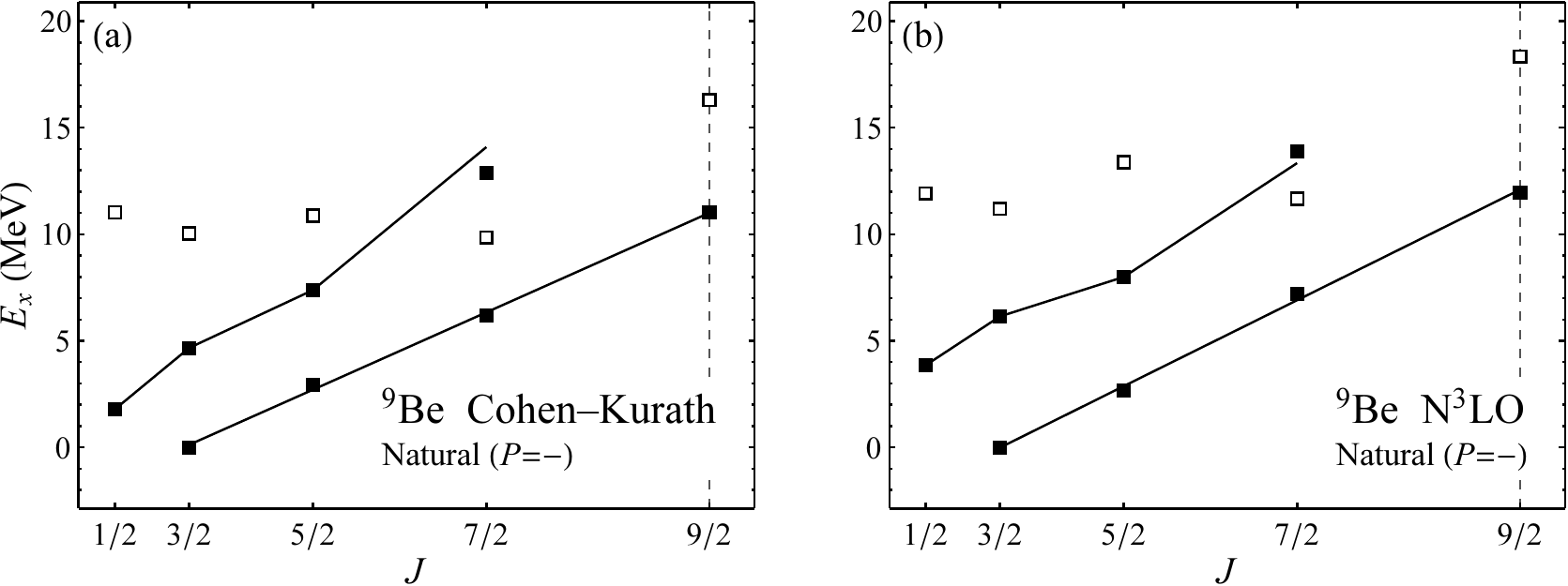}}
\caption{Excitation energies for states in  the
natural parity space of
$\isotope[9]{Be}$, for comparison with
Fig.~\ref{fig-levels-9be01}~(top),  as obtained in: (a)~a valence shell model calculation using the
Cohen-Kurath interaction and (b)~an NCCI calculation with
the \nthreelo{} interaction.
Calculations from Ref.~\protect\refcite{johnson2015:spin-orbit}.}
\label{fig-levels-9be0-johnson}      
\end{figure}

Based both on energies and transition strengths, two low-lying bands
are identified in the natural parity space, both for the JISP16
[Fig.~\ref{fig-levels-9be01}(a)] and NNLO
[Fig.~\ref{fig-levels-9be01}(b)] interactions: an yrast $K=3/2$ band
(with $J=3/2$, $5/2$, $7/2$, and $9/2$ members) and an excited $K=1/2$
band (with $J=1/2$, $3/2$, $5/2$, and $7/2$ members).  A similar
pattern of low-lying states may be found in traditional shell model
calculations with the phenomenological Cohen-Kurath $p$-shell
interaction~\cite{cohen1965:p-shell-interaction}, shown in
Fig.~\ref{fig-levels-9be0-johnson}(a).  (The maximal angular momentum
accessible within the valence space is $J=9/2$.) The apparent
restriction of these bands to the valence space (although enhanced
quadrupole transitions to off-yrast states at higher $J$ are not
excluded) and consistency with $p$-shell calculations would seem to
suggest that the structure of these bands can be largely described by
dynamics within the valence shell.  As a further indication of the
robustness of the rotational structure, similar results from NCCI calculations
with the \nthreelo{} interaction are shown in
Fig.~\ref{fig-levels-9be0-johnson}(b).
\begin{figure}[tp]
\centerline{\includegraphics[width=\hsizeforfig]{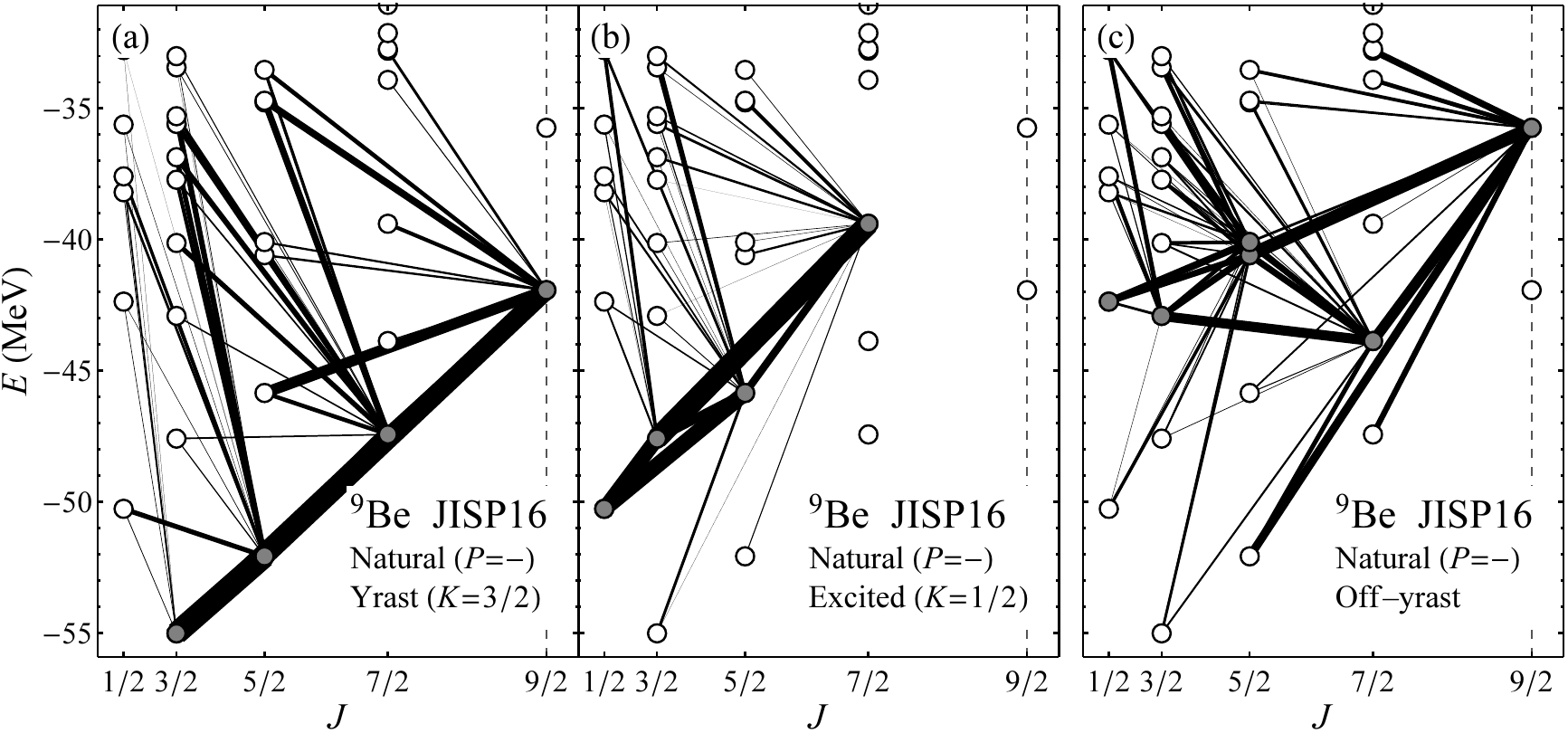}}
\caption{Electric quadrupole transition strengths among levels in the $\isotope[9]{Be}$ natural parity
  space, orignating from (a)~yrast band members, (b)~excited band
  members, and (c)~selected off-yrast states (see
  text), as obtained with the
  JISP16 nucleon-nucleon interaction.
See Fig.~\ref{fig-network-8be0} caption for discussion of plot
contents and labeling.  From
  calculations with $\Nmax=10$ at $\hw=20\,\MeV$, using the proton
  quadrupole tensor.}
\label{fig-network-9be0}      
\end{figure}

The quadrupole transition strengths from the 
band members are shown in Fig.~\ref{fig-network-9be0}(a,b).  The band
assignments are based on 
 enhanced transitions within each band.  There is also 
one example of an enhanced cross transition between the bands (namely, from the
$J=9/2$ terminating band member of the $K=3/2$ band to the $J=5/2$
member of the excited band).  

It is intriguing that the $J=7/2$ member of the excited band is the
\textit{third} $J=7/2$ state, in all four calculations considered
here, \textit{i.e.}, with JISP16 [Fig.~\ref{fig-levels-9be01}(a)],
NNLO [Fig.~\ref{fig-levels-9be01}(b)], Cohen-Kurath
[Fig.~\ref{fig-levels-9be0-johnson}(a)], and
\nthreelo{}~[Fig.~\ref{fig-levels-9be0-johnson}(b)] interactions.  The
``spectator'' $7/2_2^-$ state, rather than being part of the
rotational band structures, appears to be part of a grouping of
off-yrast states connected by enhanced quadrupole transitions, as
shown in Fig.~\ref{fig-network-9be0}(c).  This grouping is comprised also of the next
off-yrast $J=1/2$, $3/2$, $5/2$ (both members of a close doublet), and
$9/2$ states.  The W-shaped staggering pattern in the energies of these
off-yrast states (raised $1/2$, $5/2$, and $9/2$ members) is 
consistent across all four calculations.
\begin{figure}[tp]
\centerline{\includegraphics[width=\hsizeforfig]{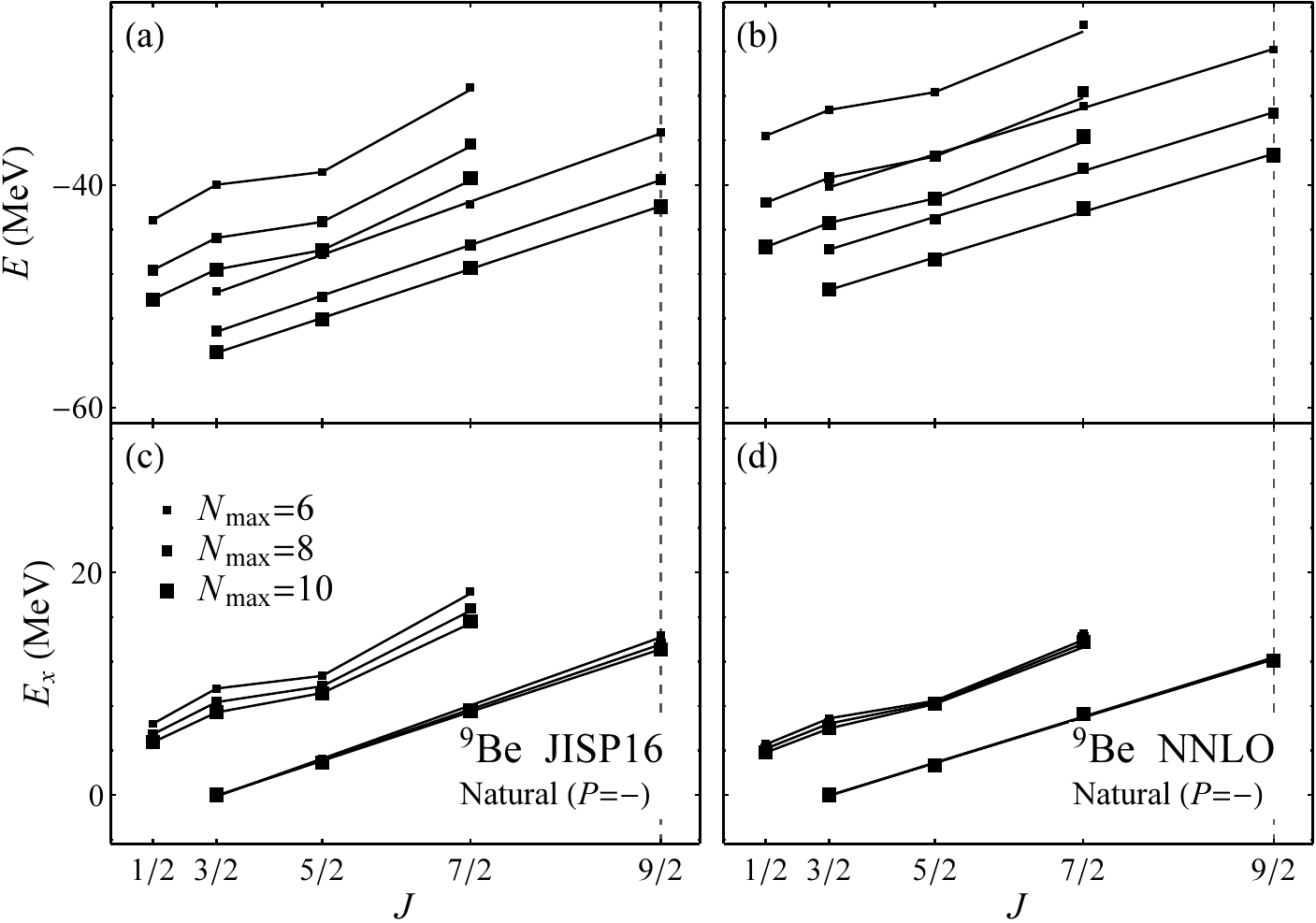}}
\caption{Dependence of calculated energies for
$\isotope[9]{Be}$ natural parity yrast and excited band members on $\Nmax$:
  energy eigenvalues $E$~(top) and excitation energies $E_x$~(bottom),
  as obtained with the JISP16~(left) and NNLO~(right)
  nucleon-nucleon interactions.  
See Fig.~\ref{fig-levels-8be0} caption for discussion of plot
contents and labeling.
From
  calculations with $\Nmax=6$, $8$, and $10$ at $\hw=20\,\MeV$.}
\label{fig-energy-Nmax-9be0}      
\end{figure}

To provide a foundation for the rotational description of
$\isotope[9]{Be}$ and to lay the groundwork for our discussion of the
band energy parameters below (Sec.~\ref{sec-be-params}), the $\Nmax$
dependence of the energies for the yrast and excited band members is
investigated in Fig.~\ref{fig-energy-Nmax-9be0}.  Much as we have
already seen for the yrast band in $\isotope[8]{Be}$
(Fig.~\ref{fig-energy-Nmax-8be0}), the calculated energy eigenvalues
shift lower by several $\MeV$ with each step in $\Nmax$
[Fig.~\ref{fig-energy-Nmax-9be0}~(top)], while the relative energies
within the band remain comparatively unchanged, as may be seen more
directly from the excitation energies
[Fig.~\ref{fig-energy-Nmax-9be0}~(bottom)].  The excitation energy of
the excited band relative to the yrast band, though not converged,
varies much less rapidly with $\Nmax$ than do the eigenvalues
themselves.

The magnetic dipole observables can provide some insight into the angular
momentum structure of a rotational band (Secs.~\ref{sec-be-8be}
and~\ref{sec-be-7be}).  However, the angular momentum structure may
also be explored more directly, by decomposing the eigenfunctions into
components of good orbital angular momentum $L$ and/or spin $S$ (which
may be further subdivided into 
proton and neutron spins, $S_p$ and $S_n$).  This decomposition is
accomplished automatically if the eigenproblem is solved from the
beginning in a basis of good orbital and spin angular momentum quantum
numbers, as in the $\grpsu{3}$-coupled NCCI code of Dytrych \textit{et
  al.}~\cite{dytrych:lsu3shell}.   However, this decomposition may
alse be extracted from wave functions obtained in conventional $M$
scheme NCCI calculations, as described in
Ref.~\refcite{johnson2015:spin-orbit} (via the so-called ``Lanczos
trick'').

\begin{figure}[tp]
\centerline{\includegraphics[width=\hsizeforfig]{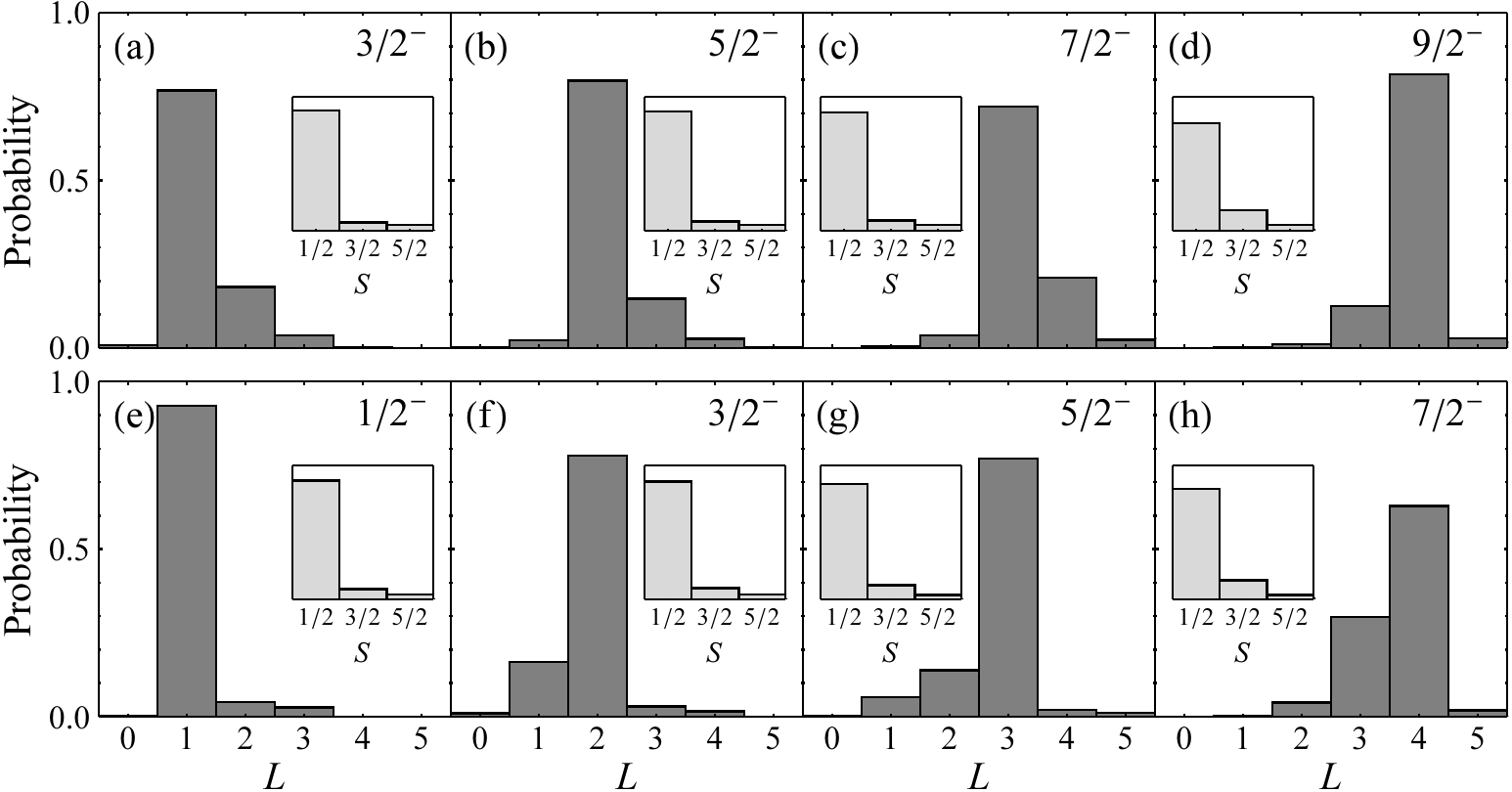}}
\caption{Decomposition of $\isotope[9]{Be}$ natural parity yrast~(top)
  and excited~(bottom) band members according to orbital angular
  momentum $L$ and spin angular momentum $S$~(insets), for the \nthreelo{} interaction.
Calculations from Ref.~\protect\refcite{johnson2015:spin-orbit}.}
\label{fig-amdecomp-9be0}      
\end{figure}

The angular momentum and spin decompositions for the natural parity
yrast and excited band members in $\isotope[9]{Be}$ are shown in
Fig.~\ref{fig-amdecomp-9be0}.  It is apparent that each state is
dominated by a specific $L$ component and by an intrinsic spin
$S=1/2$.  That is, the calculated wave functions approximately obey an
$LS$ coupling scheme~\cite{inglis1953:p-shell}.  (The angular momentum
decomposition for rotational states in $\isotope[7]{Li}$, the mirror
nucleus to $\isotope[7]{Be}$, is also explored in
Ref.~\refcite{johnson2015:spin-orbit} and yields similarly strong
evidence of $LS$ coupling.)  The $LS$ coupling, while reasonably
pronounced in Fig.~\ref{fig-amdecomp-9be0}, is not pure.  In fact, the
mixture of $L$ components in the \textit{ab initio} calculated ground
state [Fig.~\ref{fig-amdecomp-9be0}(a)] is consistent with a simple
single-irrep $\grpsu{3}$ shell-model description, including spin-orbit
interaction, as presented in Ref.~\refcite{millener2001:light-nuclei}:
an $L=1$ contribution of $21/26\approx81\%$ and an $L=2$ contribution
of $5/26\approx19\%$.

In both bands, the dominant $L$ values for successive band members are found to be
$L=1$, $2$, $3$, and $4$.  In the yrast band, the angular momenta are
coupled in the ``stretched'' (or ``aligned'') sense, with $J=L+1/2$, while,
in the excited band, these same angular momenta are coupled in the
``unstretched'' (or ``antialigned'') sense, with $J=L-1/2$.  This
pattern may be interpreted in a core-particle rotational picture, in which the core orbital
motion generates a $K=1$ band, which then couples to the neutron spin,
in aligned and antialigned senses, to generate a $K=3/2$ band and a $K=1/2$
band, respectively.  

Finally, we note that the calculated yrast states of the
\textit{unnatural} parity space of $\isotope[9]{Be}$ also constitute a
$K=1/2$ rotational band, with candidate band members at least through
$J=17/2$ [Fig.~\ref{fig-levels-9be01}~(bottom)].  The maximal angular
momentum possible in the lowest shell model unnatural parity space
(the space of $1\hw$ excitations) or, equivalently, the NCCI $\Nmax=1$
space is $J=13/2$ [dashed vertical lines in
  Fig.~\ref{fig-levels-9be01}~(bottom)].  The properties of this
unnatural parity band are similar to those already discussed for the
calculated $\isotope[7]{Be}$ yrast $K=1/2$ band (electromagnetic
observables are not shown here, but see Fig.~8 of
Ref.~\refcite{maris2015:berotor2}): (i)~band members above the maximal
valence angular momentum lie above the rotational prediction in energy
but are converging downward in energy relative to the lower band
members, (ii)~quadrupole moments and transitions are in close
agreement with the rotational predictions, except for an enhancement
in quadrupole moments relative to the rotational formula above the
maximal valence angular momentum, and (iii)~dipole moments and
transitions are generally consistent with the rotational predictions
through the highest $J$ considered, including above the maximal
valence angular momentum.


\section{Extrapolation of energies and prediction of rotational band parameters}
\label{sec-be-params}

Returning to the initial questions, from Sec.~\ref{sec-intro}, now
that we have explored how \textit{recognizable} the signatures of
rotation are, seen that they are surprisingly \textit{robust} across
interactions (and despite limitations in convergence), and inquired
into aspects of the intrinsic \textit{physical structure}, let us
touch upon how the emergent rotation compares to experiment in
quantitative detail.

\begin{figure}[tp]
\centerline{\includegraphics[width=\hsizeforfig]{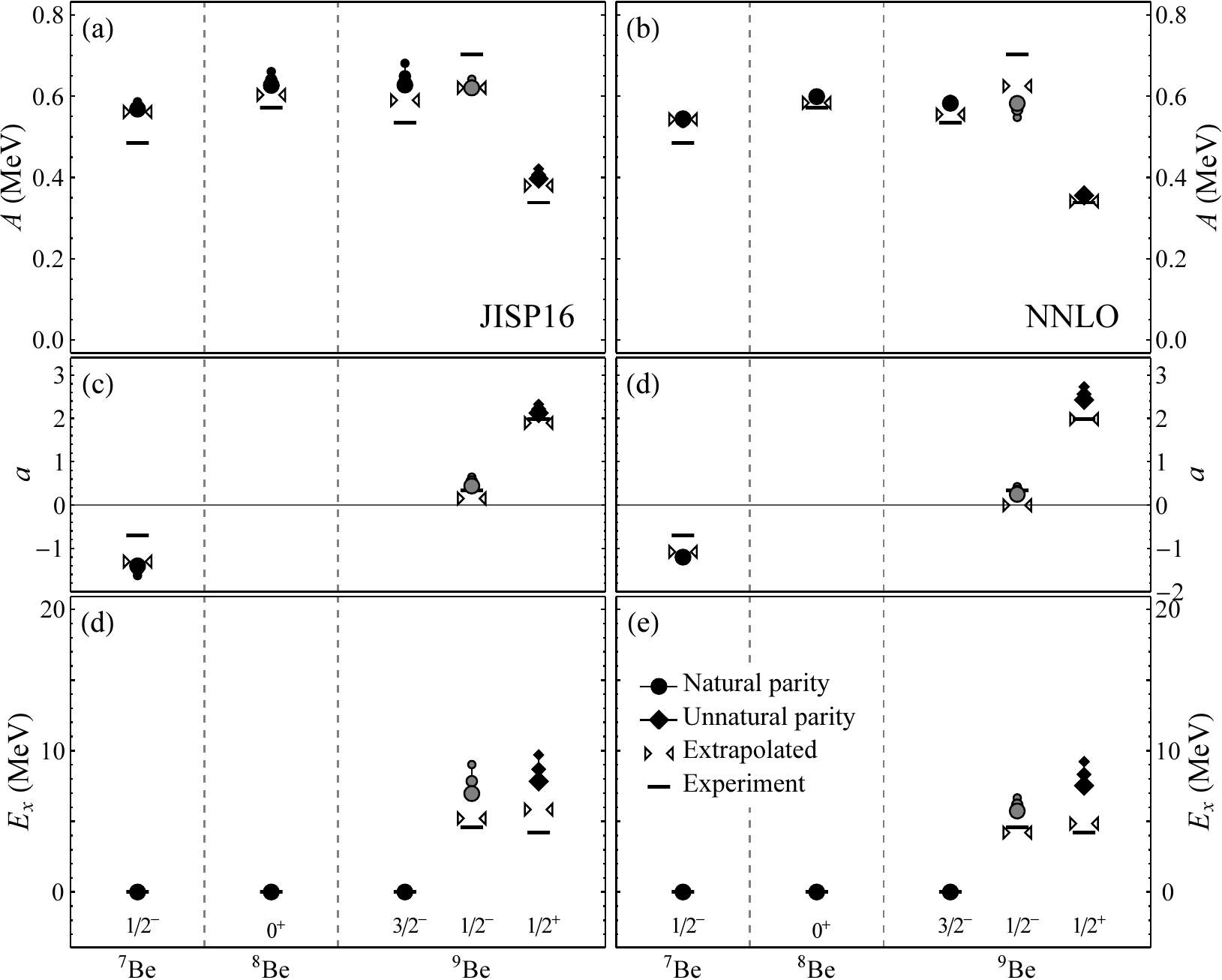}}
\caption{ Band energy parameters for $\isotope[7\text{--}9]{Be}$: the
  rotational constant $A$~(top), Coriolis decoupling parameter
  $a$~(middle), and band excitation energy $E_x$~(bottom), as obtained
  with the JISP16~(left) and NNLO~(right) nucleon-nucleon
  interactions. Values are shown for $\Nmax=6$, $8$, and $10$ (for
  natural parity) or $\Nmax=7$, $9$, and $11$ (for unnatural parity),
  with larger symbols for higher $\Nmax$ values.  Parameter values are
  also shown based on exponentially extrapolated level energies
  (paired triangles) and from the experimentally observed levels
  (horizontal lines).  }
\label{fig-params}      
\end{figure}

The energy parameters for the bands in $\isotope[7\text{--}9]{Be}$, as
extracted from level energies in the \textit{ab initio} calculations,
are summarized in Fig.~\ref{fig-params}.  Recall that the band energy
parameter $E_0$, rotational parameter $A$, and Coriolis decoupling
parameter $a$ (for $K=1/2$) entering into the rotational energy
formula~(\ref{eqn-EJ-stagger}) represent the ``height'' of the band,
the ``slope'' of the band, and the ``staggering'' of the band,
respectively, in a plot of energies \textit{vs.}\ $J(J+1)$
(\textit{e.g.}, Figs.~\ref{fig-levels-8be0}
and~\ref{fig-levels-7be0}).  The band excitation energy $E_x$, shown
in Fig.~\ref{fig-params}~(bottom), is defined relative to the yrast
band energy as $E_x\equiv E_0-E_{0,\text{yrast}}$.\footnote{Note that
  the \textit{band energy} parameter $E_0$ is the energy intercept of
  the band at $J=0$.  It is therefore \textit{not} equivalent to the
  \textit{band head energy} (except perhaps in the case of a $K=0$
  band).  Likewise, the \textit{band excitation energy}, as the
  difference of two such band energy parameters, is the vertical
  separation between the excited band and the yrast band as they
  intersect the energy axis at $J=0$.  It
  is therefore \textit{not} to be conflated with the \textit{band head
    excitation energy} (the two being equivalent only in the case
  where both bands are even-spin $K=0$ bands, thus both with $J=0$ band heads, and even
  then only in the ideal case that the band member energies lie
  exactly on the rotational line).}  Results are shown for a sequence
of $\Nmax$ truncations.  Parameters for the experimentally observed
bands (based on the set of experimental levels detailed in
Ref.~\refcite{maris2015:berotor2}) are also shown (horizontal lines).

Ideally, comparison of the calculated and experimental band
parameters provides a direct test of the degree to which the nuclear
many-body problem with the chosen internucleon interaction (here,
JISP16 or NNLO) reproduces the rotational dynamics actually occurring in the
physical $\isotope{Be}$ isotopes.  This comparison is subject 
to various challenges: the computational limitations in obtaining
convergence of energies (already discussed), the experimental challenge of identifying the
band members (in some cases from amongst broad, overlapping
resonances, with limited information available for spin-parity
assignments and almost exclusively without electromagnetic decay
data~\cite{npa2002:005-007,npa2004:008-010}), ambiguity in
describing the band through band energy parameters when level energies deviate from
the rotational formula, and the
more fundamental consideration that some of the levels involved are broad
resonances for which an equivalent sharp bound state energy is not
well-defined.  

Nonetheless, the calculated band parameters in Fig.~\ref{fig-params} are sufficiently
stable with respect to $\Nmax$ (and the experimental band parameters
sufficiently well-defined) to permit a meaningful comparison.
The experimental values of the rotational parameter $A$
[Fig.~\ref{fig-params}~(top)] for the various bands vary by about a factor
of $2$, from $\sim0.5$--$0.7\,\MeV$ for the experimental natural
parity bands down to $\sim0.34\,\MeV$ for the  
$\isotope[9]{Be}$ unnatural parity band.  The JISP16 and NNLO
calculations both consistently yield rotational parameters of $\sim0.6\,\MeV$ for the natural
parity bands and $\sim0.35$--$4\,\MeV$ for the $\isotope[9]{Be}$ unnatural parity
band. 
The Coriolis staggering for the calculated $K=1/2$ bands [Fig.~\ref{fig-params}~(middle)] varies 
in both amplitude and sign, and the experimental trend in both these properties is
reproduced across the bands.  

The excitation energies for the excited
natural parity band and the unnatural parity band in $\isotope[9]{Be}$
[Fig.~\ref{fig-params}~(bottom)]
are decreasing with $\Nmax$ [recall Fig.~\ref{fig-energy-Nmax-9be0}~(bottom)], bringing
them toward the experimental values.  
Yet, they are varying too
strongly with $\Nmax$ for it to be immediately obvious how close the
converged predictions will lie to the experimental values.

One may attempt to overcome incomplete convergence~--- and obtain more precise comparisons with the experimentally identified
rotational bands~--- by application of basis extrapolation
methods~\cite{coon2012:nscm-ho-regulator,furnstahl2012:ho-extrapolation}.
If the functional dependence (on $\Nmax$ and $\hw$) were known,
describing  how the energy eigenvalues
calculated in the
truncated spaces approach their converged values, it should,
in principle, be possible to take unconverged values
obtained from calculations in truncated spaces and use them to
estimate the true converged eigenvalues.
Such
methods are still in their formative stages.  Nonetheless, it is
intriguing to apply a straightforward scheme based on a presumed exponential convergence of energy eigenvalues
with $\Nmax$~\cite{bogner2008:ncsm-converg-2N,maris2009:ncfc}:
\begin{equation}
\label{eqn-exp}
E(\Nmax)=c_0+c_1 \exp(-c_2\Nmax).
\end{equation}
Calculations of the energy at three successive $\Nmax$ values, for fixed $\hw$, are sufficient to determine all three
parameters $c_i$ in~(\ref{eqn-exp}).  
Extrapolating to the limit $\Nmax\rightarrow\infty$ gives
 $E(\Nmax)\rightarrow c_0$, so the fitted value for $c_0$ provides an
estimate of the converged value for the energy.  The results of this
procedure are typically most stable when $\hw$ is taken near the
variational minimum in the energy curves as functions of $\hw$~\cite{maris2009:ncfc}, as in Figs.~\ref{fig-conv}(a,b), \textit{i.e.},
$\hw\approx20$--$25\,\MeV$ for the isotopes and interactions
considered here
(see
Ref.~\refcite{maris2009:ncfc} for further discussion of the procedure).
\begin{figure}[tp]
\centerline{\includegraphics[width=\hsizeforfig]{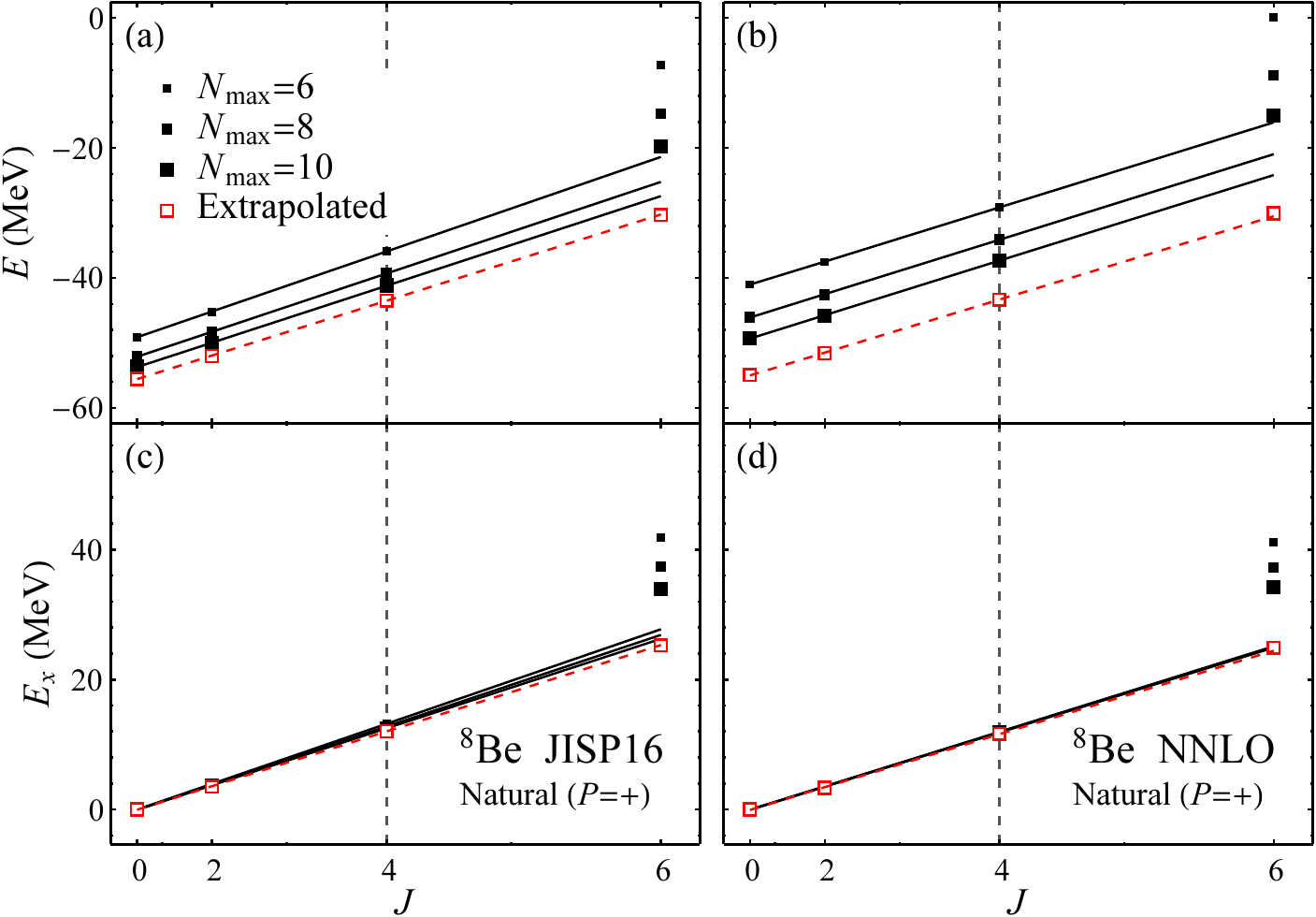}}
\caption{Calculated energy eigenvalues $E$~(top) and excitation energies $E_x$~(bottom),  for
$\isotope[8]{Be}$ natural parity yrast band members, from
  calculations with $\Nmax=6$, $8$, and $10$, as in
  Fig.~\ref{fig-levels-8be0}, together with the
  exponentially extrapolated eigenvalues~(top) and corresponding excitation energies~(bottom).}
\label{fig-energy-Nmax-8be0-exp3}      
\end{figure}

To apply this exponential extrapolation scheme to our rotational
analysis, we must first extrapolate the energies of the individual
band members.  Let us take the $\isotope[8]{Be}$ yrast band for
illustration, and revisit the $\Nmax$ dependence of the calculated
values from Fig.~\ref{fig-levels-8be0}.  The result of exponentially
extrapolating these values is seen in
Fig.~\ref{fig-energy-Nmax-8be0-exp3} (open symbols).  Observe that the
energy of the $J=6$ band member comes into line with the rotational
predictions.  Rotational energy parameters are then obtained by
matching the rotational energy formula~(\ref{eqn-EJ-stagger}) to the
extrapolated level energies (dashed lines in
Fig.~\ref{fig-energy-Nmax-8be0-exp3}).

The band energy parameters for $\isotope[7\text{--}9]{Be}$ obtained
from such extrapolations are shown in Fig.~\ref{fig-params} (paired
triangles), providing a more concrete estimate of where the converged
values lie.  Note in particular the reproduction of the excitation
energies for both excited bands in $\isotope[9]{Be}$ (at the $\MeV$
scale), as well as the reproduction of the unnatural parity band
rotational and Coriolis parameters.  This success is to be contrasted with the apparent
failure to reproduce the exceptionally high experimental rotational
parameter value of $\sim0.7\,\MeV$ for the natural parity excited
band.  However, the exponential extrapolations are subject to considerable
uncertainties~\cite{maris2009:ncfc}, and several of the experimental levels
(including in the $\isotope[9]{Be}$ natural parity excited band) are
subject to significant 
ambiguitities~\cite{npa2004:008-010,smith-etal-INPREP}.  It is
therefore not yet clear to what extent the remaining discrepancies
reflect actual deficiencies in the \textit{ab initio} description of
the nucleus with the chosen interactions, as opposed to these other limitations.


\section{Conclusions}
\label{sec-concl}

Through illustrative examples of rotational bands in \textit{ab
  initio} NCCI calculations for $\isotope[7\text{--}9]{Be}$, we have
seen how the emergence of rotational structure can be recognized
through a combination of rotational energy patterns, enhanced electric
quadrupole strengths, and general agreement of electric quadrupole and
magnetic dipole moments and transition matrix elements with rotational
predictions.  

It is simplest to recognize rotational states near the
yrast line, where the density of states is comparatively low, as in
the bands considered here.  However, rotational structure may also be
recognized in states further away from the yrast line (see the excited
$K=0$ band of
$\isotope[10]{Be}$~\cite{freer2006:10be-resonant-molecule,bohlen2008:be-band,suzuki2013:6he-alpha-10be-cluster},
as calculated and discussed in Ref.~\refcite{maris2015:berotor2}).  It is also
most straightforward to recognize rotational states with angular
momenta accessible within the traditional valence space (or NCCI
$\Nmax=0$ or $1$ space), since energies of band members above this
angular momentum may have significantly different convergence
properties and may deviate from the rotational formula, at least in the
computationally-accessible truncated calculations.

We have also begun to develop a sense of
the robustness of the emergent rotation in \textit{ab initio}
calculations: at a more fundamental level, how robustly the
imperfectly-known internucleon interaction can be expected to give
rise to rotation, and, at a more pragmatic level, how robust the
rotation is in incompletely-converged many-body calculations carried
out in truncated spaces.  

We observe a remarkable similarity in spectral details across results
obtained with two independently-derived interactions: the JISP16
interaction, obtained by inverse-scattering methods, and the chiral
NNLO interaction \nnloopt{} (calculations with a chiral \nthreelo{}
interaction were also considered).  The similarity lies not only in
the presence of analogous rotational bands across the calculations but
in the subtle deviations of these bands from the ideal rotational
formulas (\textit{e.g.}, Fig.~\ref{fig-trans-7be0}).  The similarities
arise despite the different levels and rates of convergence~---
compare, \textit{e.g.}, the energies in
Fig.~\ref{fig-energy-Nmax-8be0}(a) with those in
Fig.~\ref{fig-energy-Nmax-8be0}(b).  Comparing the predictions for
band energy parameters obtained with these two interactions, we found
a high level of quantitative consistency (Fig.~\ref{fig-params}).
Here, a simple exponential basis extrapolation scheme for energy
eigenvalues aided the comparison of the less-converged observables
(\textit{e.g.}, certain band excitation energies), but, for most of
the band energy parameters, the consistency is apparent even without
need for extrapolation.
\begin{figure}[tp]
\centerline{\includegraphics[width=\hsizeforfighalf]{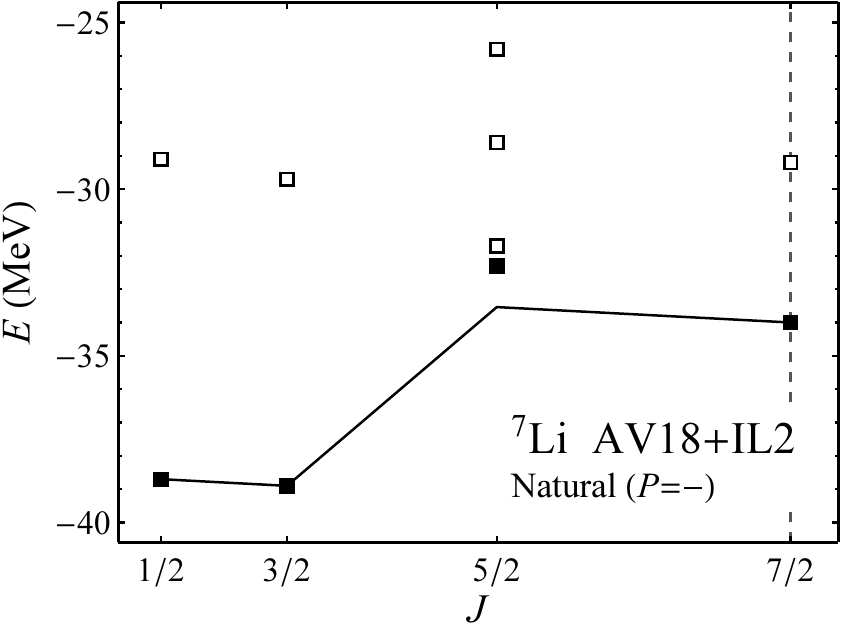}}
\caption{Energy eigenvalues for states in  the
natural parity space of
$\isotope[7]{Li}$ (the mirror nucleus to
$\isotope[7]{Be}$), as obtained in Green's Function Monte Carlo (GFMC) calculations with the AV18 nucleon-nucleon and IL2
three-nucleon interactions, for comparison with Fig.~\ref{fig-levels-7be0}.  
Calculations from Ref.~\protect\refcite{pieper2004:gfmc-a6-8}.}
\label{fig-levels-7li0-pieper}      
\end{figure}

It is particularly valuable here to note consistency across
methods.  Although the focus of this review is on NCCI calculations,
quantum Monte Carlo calculations of $\isotope[7]{Li}$ (the mirror
nucleus to $\isotope[7]{Be}$), as shown in
Fig.~\ref{fig-levels-7li0-pieper}, also readily reproduce an
$3/2$-$1/2$-$7/2$-$5/2$ yrast angular momentum
sequence~\cite{pieper2004:gfmc-a6-8}, reflective of a $K=1/2$ band
with negative Coriolis staggering.  These calculations were carried
out with an internucleon interaction consisting of an AV18 two-body part~\cite{wiringa1995:nn-av18} and an IL2 three-body
part~\cite{pieper2001:3n-il2}.  It is intriguing to note the detailed resemblance of the
eigenvalue spectrum in Fig.~\ref{fig-levels-7li0-pieper} to the NCCI
calculations of Fig.~\ref{fig-levels-7be0}.  Again the $J=5/2$ band
member lies slightly higher than the rotational formula would give,
based on the $J=1/2$, $3/2$, and $7/2$ states.  The same pattern of
off-yrast states arises, as well: the close $J=5/2$ doublet and a
subsequent set of off-yrast states ($J=1/2$, $3/2$, $5/2$, and $7/2$)
with the same staggering pattern as in Fig.~\ref{fig-levels-7be0}.

The rotational patterns are also perhaps surprisingly robust against
truncation of the many-body calculation.  The principal challenge in
identifying collective structure in NCCI calculations is the weak
convergence of many of the relevant observables (Fig.~\ref{fig-conv}).
However, there is an important distinction between convergence of
individual observables, taken singly, and convergence of relative
properties, such as excitation energies (and, especially, their
ratios) or ratios of different electromagnetic matrix elements.  It is
these latter, relative properties that are essential to the
recognition of rotation through comparison with the rotational
formula.

Our initial focus in examining observables lay simply in recognizing
rotational patterns in the \textit{ab initio} calculations and
examining their fidelity to the rotational formulas.  The existence of
such patterns suggests a \textit{rotational separation} of the wave
functions, as in~(\ref{eqn-psi}).  However, by itself, it leaves
unanswered the question of the physical origin and \textit{intrinsic
  structure} of the wave functions, \textit{i.e.}, the nature of the
intrinsic state $\tket{\phi_K}$.  
The two classic paradigms for
understanding this structure are $\alpha$ clustering and
$p$-shell dynamics, including $\grpsu{3}$ symmetry in the $p$ shell.  (A
more comprehensive, multishell framework for understanding the emergence of collective
deformation and rotational degrees of freedom is provided by $\grpsptr$
symplectic symmetry~\cite{rowe1985:micro-collective-sp6r,dytrych2008:sp-ncsm}.)

Traditionally, in rotational analysis, intrinsic matrix elements of
the multipole operators are the essential source of information on
intrinsic structure~\cite{bohr1998:v2}.  The electric quadrupole
moments and transitions provide insight into the nuclear deformation.
The absolute magnitudes of the quadrupole observables are unconverged
in the NCCI calculations (leaving only the ratio $Q_{0,p}/Q_{0,n}$,
which may provide insight into the relative deformation of the proton
and neutron distributions, as discussed in Sec.~IV\,B of
Ref.~\refcite{maris2015:berotor2}).  However, magnetic dipole
intrinsic matrix elements probe the orbital and spin angular momentum
structure of the rotational states, which may also be more directly
examined through angular momentum ($L$, $S$, $S_p$, and $S_n$)
decompositions of the wave functions.

From the examples considered here, it appears that the rotational
bands in $\isotope[7\text{--}9]{Be}$ are consistent with an
$\alpha$-$\alpha$ clustered rotational core and particle-rotor
descriptions.  Nonetheless, discontinuities in observables at the
maximal valence angular momentum suggest that the spherical shell
structure also plays some role.  It is not clear whether these
discontinuities are transient artifacts of incomplete convergence in
truncated calculations or instead persist to the full, untruncated
many-body space.  The preliminary indications vary
depending upon the observables considered, \textit{e.g.}, the
quadrupole moments in Fig.~\ref{fig-trans-8be0}~(top) or
the excitation energies in
Fig.~\ref{fig-energy-Nmax-8be0-exp3}~(bottom).

Finally, a quantitative comparison of the emergent \textit{ab initio}
rotation with experiment is subject to many challenges, from the
computational side (convergence), the experimantal side (including
identification of the relevant levels, lack of electromagnetic
transition data, and resolution of broad resonances), and the more
fundamental limitations to applying bound-state methods and a
bound-state formulation of the rotational formalism to resonant
states.  Nonetheless, despite these challenges, we have seen that the
band energy parameters extracted from \textit{ab initio} calculations
(Fig.~\ref{fig-params}) display a notable level of consistency, both
qualitative and quantitative, when compared with those extracted from
experiment.


\section*{Acknowledgements}

Discussions with M.~Freer, T.~Dytrych, S.~C.~Pieper, and R.~B.~Wiringa
are gratefully acknowledged.  We thank C.~W.~Johnson for providing
calculations from Ref.~\refcite{johnson2015:spin-orbit}.  
This material is based upon work supported by the U.S.~Department of
Energy, Office of Science, under Award Numbers~DE-FG02-95ER-40934,
DESC0008485 (SciDAC/NUCLEI), and DE-FG02-87ER40371, and by the
U.S.~National Science Foundation, under Grant Number~0904782.  An
award of computer time was provided by the U.S.~Department of Energy
Innovative and Novel Comutational Impact on Theory and Experiment
(INCITE) program.  This research used resources of the National Energy
Research Scientific Computing Center (NERSC) and the Argonne
Leadership Computing Facility (ALCF), which are DOE Office of Science
user facilities supported under Contracts~DE-AC02-05CH11231
and~DE-AC02-06CH11357. Computational resources were also provided by
the University of Notre Dame Center for Research Computing.



\end{document}